\documentclass[twocolumn]{autart} 

\newtheorem{definition}{Definition}
\newtheorem{lemma}{Lemma}
\newtheorem{remark}{Remark}
\usepackage{bbm}
\usepackage{subfigure}
\usepackage{subfigmat}
\usepackage{amsfonts}
\usepackage{algorithm}
\usepackage{algpseudocode}
\usepackage{amsmath}
\usepackage{amssymb}
\usepackage{graphicx}      
\usepackage{xcolor}
\usepackage{comment}
\usepackage{subfigure}
\usepackage{tikz,pgfplots}
\usepackage{multirow}
\pgfplotsset{compat=1.15}
\usepackage{cite}
\usepackage{color,soul}

\usepackage{ulem} 

\newcommand{\XN}{\mathbb{X}_K}

\newcommand{\X}{\mathbb{X}}
\newcommand{\U}{\mathbb{U}}
\renewcommand{\Pr}[1]{\mathsf{Pr}\left\{#1\right\}}

\def\E#1{\mathbb{E}\left\{#1\right\}}

\def\EP{\phi_\varepsilon}

\newcommand{\R}{\mathbb{R}}

\renewcommand{\S}{\mathbb{S}}

\def\R{\mathbb{R}}

\def\N{\mathbb{N}}

\def\S{\mathbb{S}}

\def\tr#1{{\rm{tr}}\left(#1\right)}
\def\cov#1{{\rm{cov}}\left(#1\right)}

\everymath{\displaystyle}
\usepackage{xcolor}

\newtheorem{proposition}{Proposition}

\newtheorem {theorem}{Theorem}
\usepackage[utf8]{inputenc}

\def\QED{\mbox{\rule[0pt]{1.3ex}{1.3ex}}} 

\def\Ec{{\mathcal{E}}}

\newcommand {\bsis} {\left\{ \begin{array} }
\newcommand {\esis} {\end{array}\right.}
 
\def\bmat#1{\left[\begin{array}{#1}}
\def\emat{\end{array}\right]}

\newcommand{\blista}{\renewcommand{\labelenumi}{(\roman{enumi})} 
	\begin{enumerate}}
	\newcommand{\elista}{\end{enumerate} \renewcommand{\labelenumi}{\arabic{enumi}.}}
\usepackage{color,soul}

\definecolor{mycolor1}{rgb}{0.00000,0.44700,0.74100}%
\definecolor{mycolor2}{rgb}{0.85000,0.32500,0.09800}%
\definecolor{mycolor3}{rgb}{0.92900,0.69400,0.12500}%

\usepackage{tikz}

\begin{document}

\begin{frontmatter}

\title{Measured-state conditioned recursive feasibility \\ for stochastic model predictive control}
%
\author[s1]{Mirko Fiacchini} \ead{mirko.fiacchini@gipsa-lab.fr}
\author[s2]{Martina Mammarella} \ead{martina.mammarella@cnr.it}
\author[s2]{Fabrizio Dabbene} \ead{fabrizio.dabbene@cnr.it}
\address[s1]{Univ. Grenoble Alpes, CNRS, Grenoble INP, GIPSA-lab, 38000 Grenoble, France.}
\address[s2]{Cnr-Istituto di Elettronica e di Ingegneria dell'Informazione e delle Telecomunicazioni, 10129 Torino, Italy.}

\begin{abstract}
In this paper, we address the problem of designing stochastic model predictive control (MPC) schemes for linear systems affected by unbounded disturbances. 
The contribution of the paper is twofold. First, motivated by the difficulty of guaranteeing recursive feasibility in this framework, due to the nonzero probability of violating chance-constraints in the case of unbounded noise, we introduce the novel definition of measured-state conditioned recursive feasibility in expectation. 
Second, we construct a stochastic MPC scheme, based on the introduction of ellipsoidal probabilistic reachable sets, which implements a closed-loop initialization strategy, i.e., the current measured-state is employed for initializing the optimization problem.
This new scheme is proven to satisfy the novel definition of recursive feasibility, and its superiority with respect to open-loop initialization schemes, arising from the fact that one never neglects the information brought by the current measurement, is shown through numerical examples. 

\end{abstract}
\end{frontmatter}

\section{INTRODUCTION}
In real-world systems and control applications, safety and performance of the system may deteriorate due to several sources of uncertainty, typically paired with the complexity of the phenomena \cite{Prekopa1995}. In the framework of constrained dynamical systems, one may resort to \textit{robust} model predictive control (MPC) schemes, implicitly or explicitly addressing worst-case realizations of the disturbance \cite{mayne2005robust}. On the other hand, some conservativeness can be reduced whenever additional information about the uncertainty is available, e.g., in the form of probability distribution, relying on \textit{stochastic} MPC (SMPC) schemes, which have proved to be the state-of-the-art control approach for uncertain systems subject to constraints that are imposed in probability, i.e., formulated as chance constraints for which a certain amount of violation is permitted (see for instance \cite{farina2016stochastic,lorenzen2016constraint, munoz2020convergence} and reference therein). 

Typically, SMPC solutions are classified as either \textit{randomized} methods, which rely on the generation of suitable disturbance realizations or scenarios \cite{Blackmore,CalafioreFagiano,lorenzen2016constraint}, or \textit{analytic approximation} methods, which exploit concentration inequalities -- as the classical Chebychev-Cantelli one -- to reformulate the probabilistic problem into a deterministic one (see \cite{farina2016stochastic, hewing2018stochastic} and references therein).

Despite the class to which the different SMPC schemes belong, all share a common issue, i.e., the inherent difficulty of guaranteeing recursive feasibility of the underlying optimization problem. Indeed, this property, which represents a key feature of classical MPC approaches, {holding also in the robust context}, becomes very hard to {be ensured} in the probabilistic framework. 
For this reason, many SMPC approaches rely on the assumption of bounded disturbance \cite{rawlings2017model} or bounded support disturbance distributions \cite{kouvaritakis2010explicit}. To this regard, it should be noted that \cite{kouvaritakis2016model} proved that probabilistic constraints on the state and/or the input can be ensured in the future only if they are satisfied for every possible realization of the uncertainty affecting the discrete-time state equation. This served as a motivation for the assumption that the support of the state equation disturbance must be assumed bounded.
In this case indeed, recursive feasibility may be ensured by robust constraint tightening techniques inspired by tube-based approaches.
In particular, when the disturbance lies in a compact set, due to the inclusion of input and state constraints, recursive feasibility can be recovered introducing a terminal cost and/or terminal constraints (see e.g., \cite{goulart2006optimization}). 

On the other hand, there are many practical situations where such boundedness assumption is just not realistic. In such cases, one has to account, by construction, for a \textit{nonzero probability} that the problem may become \textit{unfeasible}, since unbounded uncertainties almost surely lead to excursions of states from any bounded set \cite{paulson2020stochastic}. In this framework, typical solutions rely on one of the two following strategies.
The first approach relies on the definition of a \textit{backup control scheme}, which is applied whenever the system states leave the region of attraction \cite{farina2013probabilistic,paulson2020stochastic}.
In the framework of backup control strategies, a natural choice to enable the state to be steered back to the region of attraction is to soften the state constraints, exploiting a strategy similar to the \textit{exact penalty function }method \cite{di1994exact}.
In \cite{yan2018stochastic}, the chance constraints are defined as a discounted sum of violation probabilities along an infinite horizon, and by properly penalizing violation probabilities closer to the initial time while ignoring violation probabilities in the far future, the approach enables the guaranteed feasibility of the optimization problem without any assumption of boundedness of the disturbance. Similarly, in \cite{mammarella2020probabilistic} probabilistic validation techniques \cite{TeBaDa:97}, \cite{Alamo:15} have been used to guarantee recursive feasibility  without any assumption on independence or Gaussianity of the stochastic variables, through the relaxation of the chance constraints using a penalty function method \cite{kerrigan2000soft} and, following ideas presented in \cite{karg2019probabilistic}, performing an offline probabilistic design of the penalty parameter.

The second class of approaches to handle recursive feasibility in the presence of unbounded disturbance relies on a \textit{backup initialization strategy}, alternating online between a closed-loop initialization, to be chosen as long as the problem is feasible, and an open-loop one, to be adopted whenever feasibility is lost for the current observed state \cite{farina2015approach,hewing2018stochastic}. In that case, feasibility is guaranteed by utilizing the predicted state in the optimization problem,  through a proper selection of state, input, and terminal constraints. However, this choice typically worsens closed-loop performance, since it purposely neglects the information carried by the current measurements whenever the measured states are not in the region of attraction of the controller. The first attempt to properly address the problem of guaranteeing closed-loop chance constraint satisfaction and recursive feasibility was proposed in \cite{hewing2020recursively}, where an \textit{indirect feedback} over the measured state is introduced into the optimization problem through the cost function only. Then, recursive feasibility is ensured by relying on probabilistic reachable sets (PRS) \cite{pola2006invariance,hewing2018stochastic,fiacchini2021probabilistic} to design tightened constraints for the nominal system, similar to tube-based MPC \cite{rawlings2017model}.

Unlike typical open-loop initialization strategies, the peculiarity of the indirect feedback lies in the concept \textit{initial-state conditioned recursive feasibility} (ISRF). Indeed, in \cite{hewing2020recursively} the considered probabilities are \textit{conditioned} to the initial state $x_0$, and this guarantees that if the chance constraints are satisfied at time $0$, then the following \textit{realizations} of $x_k$ for any time $k>0$ would not affect the satisfaction of this constrained probability. While being practically very important, since it allows the user to be sure to have to deal always with feasible optimization problems, the ISRF concept however may easily lead to some undesirable situations. On the one hand, it may be easily happen that, even if at some instant the measured state, say $x_k=\bar x$, does violate the state constraints, the ISRF chance state constraint would still be satisfied, being conditioned only to the knowledge at time zero. On the other hand, the SMPC would be tagged as unfeasible if \textit{initialized at the same state} $\bar x$ at time $k=0$. This leads to the odd situation where the same state could be identified either as feasible or unfeasible depending on the time instant. 

The present work stems from the realization that, when dealing with unbounded stochastic uncertainty, and especially when a closed-loop initialization strategy is adopted, it becomes natural to introduce a concept of recursive feasibility using a \textit{probabilistic statement}. On the other hand, a probabilistic guarantee of feasibility is clearly not enough for our purposes, since we need to ensure that at \textit{every step} we are faced with a feasible problem, in order to be able to implement the approach.

Motivated by this consideration, we adopt a backup strategy approach, which relaxes the chance-constraints whenever the problem would not be feasible with the ``nominal" ones, but we provide probabilistic guarantees regarding these relaxation.
More specifically, we introduce a novel definition of recursive feasibility, that we name \textit{measured-state conditioned recursive feasibility (MSRF) in expectation}. A formal definition of this novel concept is provided in Section~\ref{sec:CC_approx}, but the philosophy underneath it is rather simple: we admit the constraints to be ``inflated" by a factor $\gamma$, but we guarantee that this factor will be equal to one in expectation, conditioned to the present state $x_k$ at step $k+1$ whenever the problem is feasible at step $k$, and will be decreasing in expectation otherwise.

More in details, our approach works as follows. First, similar to \cite{farina2016stochastic, hewing2018stochastic}, we recast the stochastic optimization problem into a deterministic one, where, as in tube-based approaches, the probabilistic constraint sets are tightened through \textit{ellipsoidal PRS}, which size is directly related to the desired violation level.  Importantly, different from the references above, we employ of a closed-loop initialization strategy, where at each time the deterministic problem is initialized with the current state measurement. 

Since, as commented before, the unbounded noise assumption {unavoidably leads} to non-zero probability of constraint violation, we \textit{adapt} the constraints to the \textit{current state realization} $x_k$ to avoid unfeasibility, thus obtaining scaled constraint sets that are properly relaxed whenever feasibility cannot be guaranteed for the current state. This in practice implies the adoption of a backup control strategy that steers back the state within the feasibility region, by solving a relaxed optimization problem. Specifically, we define the MSRF in expectation and a deterministic MPC formulation satisfying it, thus guaranteeing that \textit{in expectation} the values of the relaxing parameters, conditioned on the measured state $x_k$, are non-increasing over $k$. Moreover, the resulting MPC scheme leads to a convex optimization problem ensuring the satisfaction of the relaxed chance constraints at any future instant.

The paper is structured as follows. Section~\ref{sec:SMPC_intro} introduces the main SMPC framework and the novel definition of MSRF in expectation. In Section~\ref{sec:CC_approx} we introduce our main technical tools, that are the ellipsoidal chance-constraints approximations and the ensuing construction of novel ellipsoidal probabilistic reachable sets. The main technical results are presented in Section~\ref{sec:main_res}, in which we provide new SMPC schemes.
In particular, the first scheme is proven to provide guarantees of initial-state recursive feasibility when implemented with open-loop initialization. The same setup, with closed-loop initialization, is then combined in parallel with a backup recovery strategy that provides \textit{minimum} scaling relaxations, in a classical two-phases approach, that allow to recover feasibility. This scheme is proven to guarantee MSRF in expectation. Finally, the two schemes are combined, with an exact penalty approach, to provide a {compact and easily implementable SMPC scheme}, which we name {measured-state dependent SMPC} ({MS-SMPC}).
In Section~\ref{sec:pr_state} we discuss the alternative approach of relaxing the probability of satisfaction (instead of enlarging the constraint sets), and show how the two approaches are substantially dual. Input initialization strategies are discussed in Section~\ref{sec:inputbounds}. The performance of the proposed  {MS-SMPC} scheme are validated through numerical examples in Section~\ref{sec:num_sim}, where open- and closed- loop initialization approaches are compared. Main conclusions and future works are drawn in Section~\ref{sec:concl}.

{\small
\textbf{Notation.} The set $\mathbb{N}^{+}$ denotes the positive integers and $\mathbb{N} = \left\{0\right\} \cup\mathbb{N}^{+}$. 
Given $a,b\in\mathbb{N}$, $\mathbb{N}_a^{b}$
is the set of integers from $a$ to $b$.
$A\ominus B=\left\{a\in A|\,a+b\in A, \forall b\in B\right\}$ denotes the Pontryagin set difference. We use $x_k$ for the (measured) state at time $k$ and $x_{\ell|k}$ for the state predicted $\ell$ steps ahead at time $k$, to differentiate it from the realization $x_{k+\ell}$. The sequence of length $N$ of vectors $v_{0|k}, \ldots, v_{N|k}$ is denoted by $\textbf{v}_{N|k}$. The set of symmetric matrices in $\R^{n \times n}$ is denoted by $\S^n$. With $W\succ 0$
($W\succeq 0$) we denote a definite (semi-definite) positive matrix $W$. If $W\succeq 0$, then $W^{\frac{1}{2}}$ is the matrix satisfying
$W^{\frac{1}{2}} W^{\frac{1}{2}}=W$. For $W\succeq 0$, we define $\|x\|_W\doteq \sqrt{x^\top W x}$. The expected value of a random variable $x$ is denoted $\E{x}$.
Given a $W\succeq 0$ and a scalar $r \ge 0$, define $$ \Ec_{W}(r)
    \doteq
    \left\{ x=W^{1/2} z\in\R^n\,|\,  z^\top z \le r^2 
    \right\}.$$
If moreover $W\succ 0$, then
\begin{align*}
    \Ec_{W}(r)
    &=
    \left\{ x\in\R^n\,|\, x^\top W^{-1} x \le r^2 
    \right\}\\
    &=
    \left\{ x\in\R^n\,|\, \|x\|_W \le r \right\}
\end{align*}
represents the ellipsoid of ``{shape}" $W$  and ``{radius}" $r$. 
}


\section{Stochastic MPC}\label{sec:SMPC_intro}
In this section, we provide an overview on the classical definitions of the stochastic MPC problem for linear systems, and we address the problem of recursive feasibility when the system is affected by unbounded disturbances.
\subsection{SMPC problem formulation}
We consider a discrete-time, linear time-invariant system subject to additive disturbance
\begin{equation}\label{eq:sys}
x_{k+1}= A x_k + B u_k + w_k,
\end{equation}
with $x_k\in\mathbb{R}^n$, $u_k\in\mathbb{R}^m$, and $A,B$ of appropriate dimensions. The i.i.d. random noise $w_k \in \mathbb{R}^n$ is such that, for all $k \in \N$,  $\mathbb{E}\{w_k\} = 0$ and $\mathbb{E}\{w_k w_k^\top\}=\Gamma_w$. Note that, from independence, we have that $\mathbb{E}\{w_i w_j^\top\} = 0$ for all $i \neq j$. We will show later that this latter assumption may be relaxed by assuming the existence of a \textit{correlation bound} (see Remark~\ref{rem-correlation}).

Given the current state $x_k$, the states predicted $\ell$ steps ahead at time $k$, denoted as $x_{\ell|k}$, obey the following dynamics
\begin{equation}\label{eq:xell}
x_{\ell+1|k}=Ax_{\ell|k}+Bu_{\ell|k}+w_{\ell+k},\quad x_{0|k}= x_k,
\end{equation}
where $x_{\ell|k}$ are random variables (due to the random noise assumption), and the inputs $u_{\ell|k}:\mathbb{R}^n\rightarrow\mathbb{R}^m$ are assumed to be measurable functions of $x_{\ell|k}$.

We assume that the state and input are subject to polytopic chance constraints of the form
%
\begin{subequations}
\label{eq:constr}
\begin{align}
\label{eq:constr_x}
&\Pr{x \in \mathbb{X}} \geq 1-\varepsilon_x,\\
\label{eq:constr_u}
&\Pr{u \in \mathbb{U}} \geq 1-\varepsilon_u, 
\end{align}
\end{subequations}
with $\varepsilon_x,\varepsilon_u\in(0,1)$, and the polytopic sets
\begin{subequations}
\label{eq:XU}
\begin{align}
\mathbb{X} &\doteq \{ x \in \R^n\,|\, H_x x \leq h_x\}, \label{eq:X}\\
\mathbb{U} &\doteq \{ u \in \R^m\,|\, H_u u \leq h_u\}, \label{eq:U}
\end{align}
\end{subequations}
are assumed to contain the origin.
Then, the control objective is to (approximately) minimize $J_\infty$, i.e., the expected value of an infinite horizon quadratic cost
\begin{equation}
    J_\infty\doteq \lim_{j\rightarrow\infty}\mathbb{E}\Bigg\{ \frac{1}{j} \sum_{\ell=0}^{j} \big(\|x_{\ell|k}\|^2_Q +\|u_{\ell|k}\|^2_R\big)\Bigg\},
    \label{eq:cost_infty}
\end{equation}
subject to \eqref{eq:xell}--\eqref{eq:constr}, with $Q\in\mathbb{S}^{n}$, $Q\succeq 0$, $R\in\mathbb{S}^{m}$, $R\succ 0$.

To {address} this control problem, it is possible to rely on a standard SMPC approach, consisting of repeatedly solving an optimal control problem over a finite horizon~$N$, and implementing only the first control action. This approach relies on the standard assumption of the existence of an asymptotically stabilizing control gain for \eqref{eq:sys} and a suitable terminal set. 

In particular, we assume there exist a state-feedback control law $u_{k}=Kx_{k}$,
with $K\in\mathbb{R}^{m\times n}$,
and a Lyapunov matrix $P\in \mathbb{S}^{n}$, $P\succ 0$,  such that the closed-loop Lyapunov inequality  
\begin{equation}
\label{eq:LQR}
Q+K^{\top}RK+A_K^\top PA_K-P\preceq 0
\end{equation}
is satisfied, thus guaranteeing that
$A_{K} \doteq A+BK$ is Schur.
From \eqref{eq:LQR} we introduce the terminal cost $V_f(x_{N|k})=\|x_{N|k}\|_P^2$, so that the finite horizon cost $J_N(\textbf{x}_k,\textbf{u}_k)$ to be minimized at time $k$ can be defined as
\begin{equation}
    \label{eq:init_cost}
    J_N(\mathbf{x}_k,\mathbf{u}_k) \doteq\mathbb{E}\Bigg\{\sum_{\ell=0}^{N-1}\Big( \|x_{\ell|k}\|_Q^2+\|u_{\ell|k}\|^2_R\Big)+V_f(x_{N|k})\Bigg\}.
\end{equation}
Moreover, as typically done in the SMPC framework, we consider as terminal set a probabilistic invariant set\footnote{See \cite{hewing2020recursively} and the next section for a formal definition.} $\mathbb{X}_N$ contained in 
\begin{equation}
\label{eq:XN}
    \XN \doteq\Big\{x\in \mathbb{X}\,|\, 
    H_u K x\leq h_u
    \Big\},
\end{equation}
that is the set of states satisfying both state and input constraints under the stabilizing state feedback law $u_k=Kx_k$.

Similarly to what is classically done in the robust framework, see for instance \cite{mammarella2018offline}, we can split the predicted state $x_{\ell|k}$ into a deterministic, nominal part $z_{\ell|k}$, and a stochastic, error part $e_{\ell|k}$ such that
\begin{equation}
    x_{\ell|k}=z_{\ell|k}+e_{\ell|k}.
    \label{eq:nom_err}
\end{equation}
Then, we consider a prestabilizing error feedback, which leads to the following predicted input
\begin{equation}\label{eq:input}
    u_{\ell|k}=Ke_{\ell|k}+v_{\ell|k},
\end{equation}
where $K$ is the solution of \eqref{eq:LQR}, and $v_{\ell|k}$ are the free SMPC optimization variables. Hence, the nominal system and error dynamics are given {respectively} by
\begin{subequations}
\label{eq:sys_ze}
\begin{align}
z_{\ell+1|k} &= A z_{\ell|k} + B v_{\ell|k}, \quad z_{0|k}=z_{k}, \label{eq:sys_z}\\
e_{\ell+1|k} &= A_Ke_{\ell|k} + w_{k+\ell},\quad e_{0|k}=0, \label{eq:sys_e}
\end{align}
\end{subequations}
where $z_{\ell|k}$ are deterministic and $e_{\ell|k}$ are zero-mean, random variables.

To explicitly compute the expected value of $J_N(\mathbf{x}_k,\mathbf{u}_k)$ in \eqref{eq:init_cost}, it is possible to rely on \eqref{eq:sys_ze} obtaining
\begin{align}
        J_N(\mathbf{x}_k,\mathbf{u}_k) =\sum_{\ell=0}^{N-1}\Big( \|z_{\ell|k}\|_Q^2+\|v_{\ell|k}\|^2_R\Big)+V_f(z_{N|k})+c,\label{eq:cost_z}
\end{align}
where 
$$
c=\mathbb{E}\Bigg\{\sum_{\ell=0}^{N-1}\Big(\|e_{\ell|k}\|_{Q+K^\top RK}^2\Big)+V_f(e_{N|k})\Bigg\}$$
is a constant term, hence it can be neglected in the optimization. In this way, we obtain a quadratic, finite-horizon cost, which depends only on the deterministic variables $z_{\ell|k}$ and $v_{\ell|k}$.

Specifically, the chance-constrained probabilistic optimization problem 
\begin{subequations}
\label{eq:SMPC_init_S}
\begin{align}
\min_{\textbf{x}_k, \textbf{u}_k} & \mathbb{E}\Bigg\{\sum_{\ell=0}^{N-1}\Big( \|x_{\ell|k}\|_Q^2+\|u_{\ell|k}\|^2_R\Big)+V_f(x_{N|k})\Bigg\}\nonumber \\
\text{s.t.} \ \ & x_{\ell+1|k} = A x_{\ell|k} + B e_{\ell|k} + w_{k+\ell}, \\
& x_{0|k}={x_k}\\  
& \Pr{x_{\ell|k}\in\mathbb{X}}\geq 1-\varepsilon_x, \quad \ell\in\mathbb{N}_1^{N-1}\\
&  \Pr{u_{\ell|k}\in\mathbb{U}}\geq 1-\varepsilon_u,  \quad \ell\in\mathbb{N}_1^{N-1}\\
& \Pr{x_{N|k}\in\mathbb{X}_N}\geq 1-\varepsilon_N,
\end{align}
\end{subequations}
with $\varepsilon_N=\min \{\varepsilon_x,\varepsilon_u\}$, is recast into a deterministic one resorting to tube-based approaches inherited from robust MPC, i.e., 
\begin{subequations}
\label{eq:SMPC_init}
\begin{align}
\min_{\textbf{z}_k, \textbf{v}_k} &\sum_{\ell = 0}^{N-1} \left(\|z_{\ell|k}\|_Q^2+ \|v_{\ell|k}\|_R^2\right)+V_f(z_{N|k})\nonumber \\
\text{s.t.} \ \ & z_{\ell+1|k} = A z_{\ell|k} + B v_{\ell|k},\\
&z_{0|k}={x_k},\label{eq:z0-init}\\  
& z_{\ell|k}\in\mathbb{Z}_\ell, \quad \ell\in\mathbb{N}_1^{N-1}\\
& v_{\ell|k}\in\mathbb{V}_\ell, \quad \ell\in\mathbb{N}_1^{N-1}\\
& z_{N|k} \in \mathbb{Z}_N,
\end{align}
\end{subequations}
where the new tightened constraint sets $\mathbb{Z}_\ell,\,\mathbb{V}_\ell,\,\mathbb{Z}_N$ shall be properly designed to guarantee that the initial chance constraints are met in closed loop, and recursive feasibility and stability are ensured.

\subsection{On unbounded disturbances and recursive feasibility}
When the disturbances affecting the system are \textit{bounded}, it is possible to properly design the tightened constraint sets which make the optimization problem~\eqref{eq:SMPC_init} recursively feasible for $z_{0|k}=x_k$. On the other hand, in the case of \textit{unbounded} disturbances, recursive feasibility cannot in general be guaranteed for problem~\eqref{eq:SMPC_init}, and it shall be differently recovered. 

One possibility for ensuring recursive feasibility, typically employed in the framework of open-loop initialization strategy \cite{farina2015approach,hewing2018correspondence}, is to replace the measured state initialization \eqref{eq:z0-init} with $z_{0|k}=z_{1|k-1}$. That is, one considers as initial nominal state $z_{0|k}$ at time $k$ the previously computed nominal state $z_{1|k-1}$, which is in general different from the measured state $x_k$. This leads to neglect, in practice, the real value of the state at time $k$ in the SMPC control. Contextually, the nominal constraint sets can be designed 
by introducing suitable probabilistic reachable sets  and probabilistic invariant sets, which definitions are recalled hereafter.




%

\begin{definition}[Probabilistic reachable set] The sequence $\mathcal{R}_\ell$, $\ell\in\mathbb{N}$, is a sequence of probabilistic reachable sets (PRS) for the system $\xi_{\ell+1} = A\xi_\ell+w_\ell$, with violation level $\varepsilon\in[0,1)$, if the implication 
\[
\xi_0\in\mathcal{R}_0\quad \Rightarrow
\quad \Pr{\xi_\ell\in\mathcal{R}_\ell} \geq1-\varepsilon
\]
holds for all $\ell\in\mathbb{N}^{+}$.
\end{definition}
\begin{definition}[Probabilistic invariant set]
The set $\Omega \subseteq \R^n$ is a probabilistic invariant set (PIS) for the system 
$\xi_{\ell+1} = A\xi_\ell+w_\ell$, with violation level $\varepsilon \in [0,1)$,  if the implication 
\[
  \xi_0 \in \Omega \quad \Rightarrow \quad\Pr{\xi_\ell \in \Omega} \geq 1-\varepsilon
\]
holds for all $\ell\in\mathbb{N}^{+}$.
\end{definition}

In particular, it was shown in \cite{hewing2018stochastic,hewing2020recursively} that, if one selects as initial condition $z_{0|k} = z_{1|k-1}$ and designs tightened constraint sets for the nominal system \eqref{eq:sys_z} as
\begin{equation}\label{eq:ominus}
\mathbb{Z}_\ell {\subseteq} \X \ominus \mathcal{R}_{\ell}^x, \quad  \mathbb{V}_\ell {\subseteq} \U \ominus \mathcal{R}_{\ell}^u, \quad \mathbb{Z}_N\subseteq \XN \ominus \mathcal{R}_{\infty}^x,
\end{equation}
where, for a given appropriate violation level: (i) the set $\mathcal{R}_{\ell}^x$ is a $\ell$-step PRS for \eqref{eq:sys_e} with $\ell \in\mathbb{N}_0^{N-1}$; (ii) $\mathcal{R}_{\ell}^u$ is the set of $e_{\ell|k}$ such that $Ke_{\ell|k} \in \mathcal{R}_{\ell}^u$ for $\ell \in\mathbb{N}_0^{N-1}$; (iii) $\mathcal{R}_{\infty}^x$ is a PIS for \eqref{eq:sys_e}; and (iv) $\mathbb{Z}_N$ is a (properly designed) positive invariant set for \eqref{eq:sys_z} with $v_{\ell|k} = K z_{\ell|k}$. Then, the following chance-constraints
\begin{subequations}
\label{eq:constr_x0}
\begin{align}
&\Pr{x_k \in \mathbb{X} | x_0} \geq 1-\varepsilon_x, \\
&\Pr{u_k \in \mathbb{U} | x_0} \geq 1 - \varepsilon_u, 
\end{align}
\end{subequations}%
are satisfied for all $k \in \N$ with $k \ge 1$, provided that the problem \eqref{eq:SMPC_init} is feasible for $x_0$ at time $k=0$.

\begin{remark}[Initial-state conditioned RF]\label{rem:ISRF}
We refer to  condition  \eqref{eq:constr_x0} as {\rm initial-state conditioned recursive feasibility (ISRF)}, to highlight that the considered probabilities are {\rm conditioned} to the initial state $x_0$.
\end{remark}

We thus note that ISRF implies that if the chance constraints are satisfied at time $k=0$, then the following {\rm realizations} of $x_k$ for $k>0$ would not affect the satisfaction of this constrained probability. More specifically, even if at some point we verify that $x_k \not \in \X$, the ISRF chance state constraint may still be satisfied, being conditioned only to the state realization at time $k=0$. This leads to the paradoxical situation that for the same state, e.g., say $\tilde x$  not contained in $\mathbb{X}$, the SMPC might be labelled as \textit{unfeasible} if $x_0=\tilde x$ at time $k=0$, but it would be \textit{feasible} for $x_k=\tilde x$ at time $k > 0$.

For this reason, we feel the ISRF conditions \eqref{eq:constr_x0} are in spirit very different than those expressed by~\eqref{eq:constr}, which should instead be intended as being satisfied in a \textit{recurrent} way. That is, the chance-constraints should be ``updated" when the knowledge of the state at time $k$ is acquired, and the recursive feasibility shall be guaranteed accordingly. One possible way to overcome the limitations of constraining the recursive feasibility to the initial-state realization
is proposed in \cite{hewing2020recursively}. In that paper, the authors introduce an \textit{indirect feedback} over the state realizations, which enforce the information provided by the measure of $x_k$ into the deterministic optimization problem through the cost function. Still, the recursive feasibility is somehow related to Remark~\ref{rem:ISRF}. 

Another criticality inherent to the ISRF conditions is that the future predictions of the nominal state are indeed random variables when conditioned to the previous trajectory, since they are indirectly affected by the realization of the disturbance at the time of the measurement. Hence, in this paper we propose a different but effective strategy to recover recursive feasibility without neglecting the information carried by the realizations of the state, i.e.,  letting $z_{0|k}=x_k$, and considering the stochasticity of the future predicted nominal states. Indeed, it is very important to note that in general, due to the stochastic nature of the disturbance affecting the system, no guarantee can be given on the recursive feasibility of the deterministic problem \eqref{eq:SMPC_init} with initialization $z_{0|k} = x_k$, as it may inevitably lead to constraints violation. On the other hand, it would always be possible to relax the constraints on the state $\mathbb{X}$ and input $\mathbb{U}$ by introducing scaling parameters $\gamma_x(x_k),\,\gamma_u(x_k)$, and $\gamma_N(x_k)$ properly designed as functions of the realization $x_k$, to ensure the recursive feasibility of the deterministic problem and satisfaction of the closed-loop chance constraints. 

Consequently, the new deterministic optimization problem can be stated as follows\\

\begin{subequations}
\label{eq:SMPC_init_MM}
\begin{align}
\min_{\substack{\textbf{z}_k, \textbf{v}_k\\ \gamma_x, \gamma_u,\gamma_N}} &\sum_{\ell = 0}^{N-1} \left(\|z_{\ell|k}\|_Q^2+ \|v_{\ell|k}\|_R^2\right)+V_f(z_{N|k})\nonumber \\
\text{s.t.} \ \ & z_{\ell+1|k} = A z_{\ell|k} + B v_{\ell|k},\\
& z_{0|k}={x_k}\\  
& z_{\ell|k}\in\mathbb{Z}_\ell\subseteq\gamma_x\mathbb{X}\ominus\mathcal{R}^x_{\ell}, \quad \ell\in\mathbb{N}_1^{N-1}\\
& v_{\ell|k}\in\mathbb{V}_\ell\subseteq\gamma_u\mathbb{U}\ominus\mathcal{R}^u_{\ell}, \quad \ell\in\mathbb{N}_1^{N-1}\\
& {z_{N|k} \in \mathbb{Z}_N \subseteq \gamma_N \XN \ominus \mathcal{R}^x_{\infty} } \\
& {\gamma_x \geq 1, \quad \gamma_u \geq 1, \quad \gamma_N \geq 1.}
\end{align}
\end{subequations}
Then, denoting as $\gamma_x^\star(x_{k}), \, \gamma_u^\star(x_{k})$, and $\gamma_N^\star(x_{k})$ the solutions of \eqref{eq:SMPC_init_MM} at time $k$, the feasibility of the problem at the next step $k+1$ might lead to different scaling parameters $\gamma_x^\star(x_{k+1}), \,\gamma_u^\star(x_{k+1}),\,\gamma_N^\star(x_{k+1})$ depending on the measured state $x_{k+1}$. Hence, the scaling parameters at time $k+1$ cannot be determined at time $k$. Nevertheless, their prior value can be modelled  at time $k$ by random variables, thus allowing one to infer recursive feasibility properties to Problem~\eqref{eq:SMPC_init_MM} in terms of the expectations of its solutions at time $k+1$, conditioned to the knowledge at $k$.

The main aim of this paper is to define a novel \textit{probabilistic property on the scaling parameters}, and a deterministic MPC formulation satisfying it, thus guaranteeing that in expectation their values, conditioned on the measured state $x_k$, are non-increasing over $k$. Moreover, the resulting MPC scheme should ensure satisfaction of the relaxed chance constraints at any future instant. We introduce this new notion of recursive feasibility, i.e., the \textit{measured-state conditioned recursive feasibility in expectation}, in the following definition.

%


\begin{definition}[MSRF in expectation]
\label{def-MSRF}
The optimiza\-tion problem \eqref{eq:SMPC_init_MM} is said to satisfy the \textrm{measured-state conditioned recursive feasibility in expectation} at time $k \in \N$ if its optimal solutions 
$\gamma_x^\star(x_k)$, $\gamma_u^\star(x_k)$, and $\gamma_N^\star(x_k)$ are such that
\begin{subequations}
\label{eq:RCFE_def}
\begin{align}
\mathbb{E}\left\{ \gamma_x^\star(x_{k+1})| \ x_k \right\} & \le \gamma_x^\star(x_{k}),\\
\mathbb{E}\left\{ \gamma_u^\star(x_{k+1})| \ x_k \right\} & \le \gamma_u^\star(x_{k}),\\
\mathbb{E}\left\{ \gamma_N^\star(x_{k+1})| \ x_k \right\} & \le \gamma_N^\star(x_{k}),
\end{align}
\end{subequations}
and the chance constraints conditioned to $x_k$, i.e., 
\begin{subequations}
\label{eq:constr_xk}
\begin{align}
&\Pr{x_{\ell|k}\in \gamma_x^\star(x_{k})\mathbb{X} | \ x_k} \geq 1-\varepsilon_x,\label{eq:constr_xkx}\\
&\Pr{u_{\ell|k}\in \gamma_u^\star(x_{k})\mathbb{U} | \ x_k} \geq 1 - \varepsilon_u,\label{eq:constr_xku}
\end{align} 
\end{subequations}
hold for all $\ell \in \N$.
\end{definition}

Clearly, if no scaling is required for \eqref{eq:SMPC_init_MM} to be feasible at $k$, i.e., $\gamma_x^\star(x_k) = \gamma_u^\star(x_k) = \gamma_N^\star(x_k) = 1$, then
\begin{align*}
&\mathbb{E}\left\{\gamma_x^\star(x_{k+1})| \ x_k \right\} = 1, \quad \mathbb{E}\left\{\gamma_u^\star(x_{k}+1)| \ x_k \right\}=1,\\
&\hspace{2cm} \mathbb{E}\left\{\gamma_N^\star(x_{k+1})| \ x_k \right\} = 1,
\end{align*}
meaning that no relaxation will be needed in expectation. Moreover, it is worth to highlight that if $\gamma_x^\star(x_0) = \gamma_u^\star(x_0) = \gamma_N^\star(x_0) = 1$ at time $k=0$, then the initial-state conditioned recursive feasibility is ensured for Problem~\eqref{eq:SMPC_init_MM}.

Based on this novel concept, in the following we propose a SMPC structure guaranteeing MSRF in expectation of problem \eqref{eq:SMPC_init_MM} and such that the chance constraints \eqref{eq:constr_xkx}--\eqref{eq:constr_xku} are satisfied by the future realization of the system. Specifically, we will present an approach to compute ellipsoidal PRS based on the knowledge of the mean and the variance of the disturbance, which will allow to construct appropriate ellipsoidal tightening sets, guaranteeing closed-loop constraints satisfaction and recursive feasibility for problem \eqref{eq:SMPC_init_MM}. Moreover, we will employ a penalization on the scaling parameters to minimize the constraints relaxation, avoiding it if possible. Finally, we will formulate the SMPC problem as a convex optimization problem in order to guarantee optimality, measured-state conditioned RF in expectation, closed-loop chance constraints satisfaction, and convergence of the solution to the origin. 

\section{Chance constraints ellipsoidal tightening}
\label{sec:CC_approx}
Let us consider the error dynamics in \eqref{eq:sys_e}. Following the assumptions on the disturbance $w_{k+\ell}$, we have $\E{e_{\ell|k}} = 0$ while the covariance matrix $E_{\ell|k}\doteq \mathbb{E}\{e_{\ell|k} e_{\ell|k}^\top\}$ satisfies the following recursion
\begin{equation}
E_{\ell+1|k} = A_K E_{\ell|k} A_K^\top + \Gamma_w, \label{eq:Ek}
\end{equation}
for all $k\in\mathbb{N}_0$, with $E_{0|k} = 0$.

Then, to enforce the probabilistic constraints, we adopt an approach based on the classical Chebychev inequality. In particular, we observe that, as discussed in \cite{kofman2012probabilistic,hewing2018correspondence,fiacchini2021probabilistic},  {the Chebychev inequality with the covariance matrix recursion provided in \eqref{eq:Ek} can be exploited to derive a sequence of ellipsoids that are guaranteed to contain the error $e_{\ell|k}$ with desired probability.} More precisely, we have that the ellipsoid $\Ec_{E_{\ell|k}}\!(\rho)$ with shape matrix $E_{\ell|k}$ and radius $\rho$ satisfies the following chance constraint
\begin{equation}
\label{Cheby}
{\Pr{e_{\ell|k}\in\Ec_{E_{\ell|k}}\!(\rho)}} \geq 1 - \frac{n}{\rho^2},
\end{equation}
where the probability in \eqref{Cheby} is measured according to a generic distribution of $w_k$.
In the case {$e_{\ell|k}$} follows a {Normal} distribution, we can obtain a better bound, given by
\begin{equation}
{\Pr{e_{\ell|k}\in\Ec_{E_{\ell|k}}\!(\rho)}} \geq\chi_{n}^2(\rho^2).
\end{equation}
To simplify the subsequent developments, we introduce the violation level function 
\begin{equation}
    \label{eq:eps_r}
\EP(\rho)  \doteq 
\begin{cases}
   \frac{n}{\rho^2} &\text{ for generic distributions},\\
  1 - \chi_n^2(\rho^2) &\text{ for Normal distributions} ,
\end{cases}
\end{equation}
and we obtain the general bound
\begin{equation}
\label{eq:prob_phi}
{\Pr{e_{\ell|k}\in\Ec_{E_{\ell|k}}\!(\rho)}} \geq 1- \EP(\rho).
\end{equation}
Note that $\EP(\rho)$ is a strictly decreasing function of~$\rho$, hence the smaller $\rho$ is, i.e.,\ the larger the violation level is, the smaller the probabilistic guarantee will be. This is reasonable, since smaller $\rho$ corresponds to smaller ellipsoid $\Ec_{E_{\ell|k}}\!(\rho)$.

\subsection{Construction of ellipsoidal PRSs}
We exploit the previous result to construct a sequence of probabilistic reachable sets for the error dynamics \eqref{eq:sys_e} subject to \textit{uncorrelated} noise on a $N$-step prediction horizon. To this end, we can rely on a method similar to the one proposed in \cite{fiacchini2021probabilistic}. In particular, we exploit \cite[Proposition 4]{fiacchini2021probabilistic}, which states that, if we can find a matrix $W_x\succ 0$ and a parameter $\lambda \in (0, 1)$ such that 
\begin{subequations}
\label{eq:reachLMI}
    \begin{align}
 A_K W_x A_K^\top &\preceq \lambda^2 W_x  , \label{eq:reachLMI_1}\\
 \Gamma_w  &\preceq (1 - \lambda)^2 W_x, \label{eq:reachLMI_2}
\end{align}
\end{subequations}
then, for $\rho>0$  the sequence of ellipsoids of shape $W_x$ and increasing radius $\rho( 1-\lambda^\ell)$, i.e.,
\begin{equation}\label{eq:PRS_ell}
\mathcal{R}_\ell^x = \Ec_{W_x}\!(\rho(1-\lambda^\ell)),\quad {\ell\in\mathbb{N}},
\end{equation} 
represents a sequence of PRS for \eqref{eq:sys_e} with violation probability $\EP(\rho)$. Formally, {for every $\ell\in\mathbb{N}$},  we have 
\begin{equation}\label{eq:Pre}
    \Pr {e_{\ell|k}\in \Ec_{W_x}\!(\rho(1-\lambda^\ell))}\geq 1-\EP(\rho),
\end{equation}
with $\EP(\rho)$ defined as in \eqref{eq:eps_r}.

\begin{remark}[On correlated noise]
\label{rem-correlation}
To ease the analysis, following the majority of the works concerning probabilistic reachable and invariant sets computation and SMPC, we modelled the stochastic disturbance $w_k$ by a sequence of i.i.d. random variables. However, requiring constant mean and constant covariance matrices of the disturbance, and their exact knowledge, may be a too restrictive assumption in practice, when dealing with real systems and data. To this regard, we remark that we may resort on the techniques introduced in \cite{fiacchini2021probabilistic}, where the authors introduce the so-called \textit{correlation bound} for systems excited by disturbances whose realizations are correlated in time and whose moments are only partially known. In particular, it was proven in \cite{fiacchini2021probabilistic} that the existence of such correlation bounds requires only the existence of bounds on the mean and the covariance matrices, and a Schur stability condition on $A_K$.   
\end{remark}

\subsection{Ellipsoidal approximations of constraint sets}

Let us consider the polytopic set $\mathbb{X}$ defined in \eqref{eq:X}, and let $W_x$ be solution of \eqref{eq:reachLMI} for some $\lambda\in (0,1)$. Then, we compute the maximum radius $r_x$ of the ellipsoid of shape $W_x$ inscribed in $\mathbb{X}$, solving the convex optimization problem
\begin{align}
r_x \doteq \arg&\max p \label{eq:rx} \\ 
&\text{s.t. } \Ec_{W_x}\!(p) \subseteq\mathbb{X}.\nonumber
\end{align}
Clearly, the ensuing ellipsoid is such that 
\begin{equation}\label{eq:Wx}
\Ec_{W_x}\!(r_x) = \{x \in \R^{n}: \ x^\top W_x^{-1} x \leq r_x^2\} \subseteq \X. 
\end{equation}
Note that the possible choice of parameters $K$, $W_x$, $\lambda$ and $r_x$ is not unique, as there are multiple objectives. One, in fact, may want to define the parameters to have fast convergence for the nominal state, making $\lambda$ small, but also to obtain a set $\Ec_{W_x}\!(r_x)$ providing a good approximation of $\X$. On the other hand, a small $\lambda$ might lead to an aggressive local control $K$, and then to too tight input constraints on $v_{\ell|k}$. With conditions \eqref{eq:reachLMI} and \eqref{eq:rx} we propose one possibility, consisting in first fixing $K$, $W_x$ and $\lambda$ to have the desired nominal contraction $\lambda$, and then scaling the set defined by $W_x$ to better fit in $\X$.

Regarding the input constraint set $\mathbb{U}$ in \eqref{eq:U}, we can proceed in an analogous way. In particular, we first fix an (arbitrary) positive value $\hat{r}_u>0$, and compute the largest ellipsoid $\Ec_{\hat{W}_u}(\hat{r}_u)$ contained in $\U$ as the solution of the following optimization problem
\begin{align*}
\hat{W}_u \doteq \arg &\max_{W\succ 0} \log\det W\\ 
&\text{s.t. } \Ec_{W}(\hat{r}_u) \subseteq \U \nonumber.
\end{align*}

Then, to construct the terminal constraint set contained in $\XN$, ensuring the desired invariance properties, the matrix $\hat W_u$ is must be appropriately scaled. 
To this end, we note that we can always rescale the matrix $\hat W_u$ as $W_u=\eta \hat W_u$ so that the following inequality 
\begin{equation}\label{eq:WuWx}
K^\top W_u^{-1} K \preceq W_x^{-1}
\end{equation}
holds.
In particular, this can be obtained by solving the following convex problem
\begin{eqnarray*}
\eta^\star &= \arg\min & \eta\\
&\text{ s.t. } &K^\top \hat{W}_u^{-1} K \preceq \eta W_x^{-1} \nonumber
\end{eqnarray*}
and defining $W_u = \eta^\star \hat{W}_u$ and $r_u = \hat{r}_u/\sqrt{\eta^\star}$.
Note that the set $\Ec_{W_u}(r_u)$ is the maximal-volume ellipsoid whose elements satisfy the input constraints, i.e.,
\begin{equation}\label{eq:Wu}
\Ec_{W_u}(r_u) = \{u \in \R^{m}: \ u^\top W_u^{-1} u \leq r_u^2\} \subseteq \U, 
\end{equation}
and, moreover, \eqref{eq:WuWx} guarantees that $x \in \Ec_{W_x}(r)$ implies $Kx \in \Ec_{W_u}(r)$ for all $r \geq 0$, and then the set $\mathcal{R}_\ell^u$ in \eqref{eq:ominus} can be given by
\begin{equation}\label{eq:PRS_ell_u}
\mathcal{R}_\ell^u = \Ec_{W_u}(\rho (1-\lambda^\ell)),  \quad \forall \ell \in \N.
\end{equation}

\begin{lemma}\label{lem:rW}
Let matrices $W_x, W_u$ and radii $r_x, r_u$ satisfy \eqref{eq:Wx}, \eqref{eq:Wu}, and \eqref{eq:WuWx}. Then 
\[
\Ec_{W_x}(r_N) \subseteq \XN,
\quad r_N \doteq \min\{r_x, \, r_u\},
\]
that is, if $x \in \Ec_{W_x}(r_N)$, then $x \in \Ec_{W_x}(r_{{N}}) \subseteq \X$ and $u = Kx \in \Ec_{W_u}(r_{{N}}) \subseteq \U$.  
\end{lemma}

\begin{proof}
Recall that, from \eqref{eq:WuWx}, $x \in \Ec_{W_x}(r_N)$ implies $Kx \in \Ec_{W_u}(r_N)$.  Consider first the case $r_N = r_x \leq r_u$. Then, $x \in \Ec_{W_x}(r_N) = \Ec_{W_x}(r_x) \subseteq \X$ and $u = Kx \in \Ec_{W_u}(r_N) \subseteq \Ec_{W_u}(r_u) \subseteq \U$. Analogously, if $r_N = r_u < r_x$, $x \in \Ec_{W_x}(r_N) \subset \Ec_{W_x}(r_x) \subseteq \X$ and $u = Kx \in \Ec_{W_u}(r_N) = \Ec_{W_u}(r_u) \subseteq \U$. 
\end{proof}

The derivations presented above allowed us to construct proper ellipsoidal inner approximations $\Ec_{W_x}\!(r_x)\subseteq \X$,  $\Ec_{W_u}(r_u)\subseteq \U$, and $\Ec_{W_x}\!(r_N)\subseteq \XN$. To this regard, the following remark guarantees that any scaling of the original constraint sets are reflected in their corresponding ellipsoidal approximations.

\begin{remark}[Scaling of  ellipsoid approximations]
We remark  that any scaling of the sets $\X$ and $\U$ would lead to the same scaling for their inner approximating ellipsoids $\Ec_{W_x}(r_x)$ and $\Ec_{W_u}(r_u)$, that is 
\begin{align*}
\Ec_{W_x}(\gamma r_x) = \gamma \Ec_{W_x}(r_x) \subseteq \gamma \X,\\
\Ec_{W_u}(\gamma r_u) = \gamma \Ec_{W_u}(r_u) \subseteq \gamma \U.
\end{align*}
for all $\gamma\ge 0$. Thus, any relaxation on the ellipsoids shall be seen as a relaxation on the state and input polytopes.
\end{remark}

\subsection{Ellipsoidal tightening of state and input constraints}
\label{sec:tight}

Let us now focus on the nominal system \eqref{eq:sys_z}, and consider the stabilizing control law $v_{\ell| k}= K z_{\ell|k}$. Then, if we select $W_x\succ 0$ and $\lambda \in (0,1)$ to satisfy \eqref{eq:reachLMI_1}, simple derivations show that $W_x$ also satisfies
\begin{equation}\label{eq:lambda}
A_K^\top W_x^{-1} A_K \preceq \lambda^2 W_x^{-1}.
\end{equation}
In turn, this implies that, in the absence of additive disturbance, the quadratic function associated to $W_x^{-1}$ is \textit{exponentially decreasing} with convergence rate $\lambda$ along the trajectory of the nominal system in closed-loop with $v_{\ell|k} = K z_{\ell|k}$. Then, we define with $\Ec_{W_x}\!(\alpha_\ell)$ the ellipsoid of shape $W_x$ having $z_{\ell|k}$ on its boundary. This can be obtained letting
$\alpha_\ell = \Big(z_{\ell|k}^\top W_x^{-1}z_{\ell|k}\Big)^{1/2}$.
Hence, we have that $z_{\ell|k} \in \Ec_{W_x}\!(\alpha_\ell)$ and
\begin{equation}\label{eq:convergence}
z_{\ell+j|k} = A_K^j z_{\ell|k} \in \Ec_{W_x}\!(\alpha_\ell \lambda^j) 
\end{equation}  
for all $j \in \N_0^{N-1}$. 

Now, with the following proposition we outline a condition on the nominal predicted states $z_{\ell|k}$ to guarantee that the chance constraints \eqref{eq:constr_xkx} hold for $\ell \in \N$. 

\begin{proposition}[Tightened state constraints]
\label{prop:xellk}
Given the system \eqref{eq:sys_ze}, $W_x\succ 0$ and $\lambda \in (0,1)$ satisfying \eqref{eq:reachLMI}, and $r_x$ as in \eqref{eq:rx}.
For each $\ell \in \N$, let $\rho>0$ be such that $\rho(1-\lambda^\ell)\le r_x$. If  $z_{\ell|k}$  satisfies
\begin{align}
\label{eq:prop1}
z_{\ell|k}&\in
\Ec_{W_x}\!\left(r_x-\rho(1-\lambda^\ell) \right)\\
&\quad =\Big\{
z_{\ell|k} \,:\, 
z_{\ell|k}^\top W_x^{-1}z_{\ell|k}\leq \Big(r_x-\rho(1-\lambda^\ell)\Big)^2
\Big\}, \nonumber
\end{align}
then the recurrent state chance-constraints
\begin{equation}\label{eq:Prop1a}
    \Pr{x_{\ell|k}\in \mathbb{X} | x_k} \geq 1- \EP(\rho), 
\end{equation} 
are satisfied for $\ell \in \N$. 
\end{proposition}

\begin{proof}
First, notice that, by definition, $z_{\ell|k} \in \Ec_{W_x}\!(\alpha_\ell)$ with $\alpha_\ell = \Big(z_{\ell|k}^\top W_x^{-1}z_{\ell|k}\Big)^{1/2}$. Combining  \eqref{eq:ominus} and \eqref{eq:PRS_ell}, one obtains
\begin{equation}\label{eq:prop1b}
     \Pr{x_{\ell|k} \in \Ec_{W_x}\Big(\alpha_\ell+\rho(1-\lambda^\ell)\Big)}\geq 1-\EP(\rho).
 \end{equation}
Then, to ensure \eqref{eq:Prop1a}, being $\Ec_{W_x}\!(r_x)  \subseteq\mathbb{X}$ by construction, it is sufficient to prove the following set inclusion
\begin{equation}
   \Ec_{W_x}\Big( \alpha_\ell+\rho(1-\lambda^\ell)\Big)  \subseteq \Ec_{W_x}\!(r_x),
\end{equation}
which holds if and only if $\alpha_\ell+\rho(1-\lambda^\ell) \leq r_x$ or, equivalently, if 
\begin{equation*}
z_{\ell|k}^\top W_x^{-1}z_{\ell|k}\leq \Big(r_x-\rho(1-\lambda^\ell)\Big)^2,
\end{equation*}
that is if \eqref{eq:prop1} is satisfied.
\end{proof}

Similarly, the following proposition provides the analogous conditions for the nominal input to guarantee chance constraints satisfaction along the horizon.
 
\begin{proposition}[Tightened input constraints]\label{prop:xellkuellk}
Given the system \eqref{eq:sys_ze}, $W_x\succ 0$ and $\lambda \in (0,1)$ satisfying \eqref{eq:reachLMI}, and $W_u\succ 0$ and $r_u>0$ such that \eqref{eq:Wu} holds. For each $\ell \in \N$, let $\rho>0$ be such that $\rho (1-\lambda^\ell) \le r_u$. If $v_{\ell|k}$
satisfies
\begin{align}
\label{eq:prop2}
v_{\ell|k}&\in
\Ec_{W_u}\!\big(r_u-\rho (1-\lambda^\ell)\big)\\
&\quad=
\Big\{
v_{\ell|k} \,:\, 
v_{\ell|k}^\top W_u^{-1} v_{\ell|k}\leq (r_{u} - \rho (1-\lambda^\ell))^2\Big\},\nonumber
\end{align}
then the recurrent input chance-constraints
\begin{equation}
    \Pr{u_{\ell|k}\in \mathbb{U} | \, x_k} \geq 1 - \EP(\rho).
\end{equation}
 are satisfied
for $\ell \in \N$.  
\end{proposition}

\begin{proof} Consider the constraints on the nominal input $v_{\ell|k}$. By construction, see \eqref{eq:WuWx}, if $e_{\ell|k} \in \Ec_{W_x}\!\left(\rho(1-\lambda^\ell)\right)$ then $K e_{\ell|k} \in \mathcal{R}_\ell^u = \Ec_{W_u}(\rho (1-\lambda^\ell))$. Since $e_{\ell|k} \in \Ec_{W_x}\!\left(\rho(1-\lambda^\ell)\right)$ holds with probability $1-\EP(\rho)$ by definition, being the set  $\Ec_{W_x}\!\left(\rho(1-\lambda^\ell)\right)$ the PRS at time $\ell$ for \eqref{eq:sys_e}, it follows that 
\begin{equation}\label{eq:prop2a}
v_{\ell|k} \in \Ec_{W_u}(r_u - \rho (1-\lambda^\ell))
\end{equation}
implies $u_{\ell|k} = v_{\ell|k} + K e_{\ell|k} \in \Ec_{W_u}(r_u)$ with the same probability $1-\EP(\rho)$. Hence, the input chance constraints are satisfied with probability no smaller than $1-\EP(\rho)$ if 
\begin{equation*}
v_{\ell|k}^\top W_u^{-1} v_{\ell|k}\leq (r_{u}- \rho (1-\lambda^\ell))^2.
\vspace*{-0.5cm}
\end{equation*}
\end{proof}

\begin{remark}
[On violation probabilities and $\rho$]
We  observe that Propositions~\ref{prop:xellk} and~\ref{prop:xellkuellk} provide, respectively, convex (ellipsoidal) conditions on $z_{\ell|k}$ and $v_{\ell|k}$ to ensure that the state and input chance constraints are satisfied for $\ell \in \N$ with a desired bound on the violation level. In particular, one can directly control the level of violation acting on the radius $\rho$. In fact, when we need to satisfy different violation levels $\varepsilon_x,\varepsilon_u$
for the state and input, as in \eqref{eq:constr_xk}, one can just select radii $\rho_x=\EP^{-1}(\varepsilon_x)$ 
and 
$\rho_u=\EP^{-1}(\varepsilon_u)$
in \eqref{eq:prop1} and \eqref{eq:prop2}, respectively. Moreover, we could easily handle time-varying chance-constraints, for instance assuming some dependence of $\varepsilon_x$ from $\ell$. In this case, it would suffice to select a time-varying $\rho(\ell)$.
\end{remark}

Once this interchangeability interpretation between the radius $\rho$ and the violation level $\varepsilon$ is clarified, in the sequel we will make the following simplifying assumption 
\begin{equation}
  \varepsilon_x=\varepsilon_u=\varepsilon.
\end{equation}
This will allow us to ease the derivations, without loosing the main message we want to convey. 
In particular, we can define the tightened state and input constraint as follows
\begin{align}
\label{eq:Zl}
\mathbb{Z}_\ell & = \Ec_{W_x}\!\big(r_x-\rho(1-\lambda^\ell)\big),\\
\label{eq:Vl}
\mathbb{V}_\ell & = \Ec_{W_u}\big(r_u-\rho (1-\lambda^\ell)\big).
\end{align}

\subsection{Tightened terminal constraint set}
To design the terminal constraint set, it is necessary to search for a condition on the nominal state at time $\ell = N$ that implies the chance constraints satisfaction at time~$N$ and also in the whole future. To do this, we rely again on the concepts of PRS and PIS, as shown in the next theorem. 

\begin{theorem}\label{th:term}
Given the system \eqref{eq:sys_ze}, $W_x\succ 0$ and $\lambda \in (0,1)$ satisfying \eqref{eq:reachLMI}, and $r_N= \min\{r_x, r_u\}$ such that \eqref{eq:Wx}, \eqref{eq:Wu}  and \eqref{eq:WuWx} hold, and $\rho \leq r_{N}$. If $z_{N|k}$ satisfies 
\begin{align}
\label{eq:constr_zNk}
z_{N|k}&\in
\Ec_{W_x}\!\left(r_{N} - \rho(1-\lambda^N) \right)\\
&\quad =
\Big\{
z \,:\, 
z^\top W_x^{-1} z \leq \Big(r_N-\rho(1-\lambda^\ell)\Big)^2
\Big\}, \nonumber
\end{align}
then the terminal chance constraint 
\begin{align}
&\Pr{x_{\ell|k}\in \XN | x_k} \geq 1-\EP(\rho),\label{eq:constr_xkXN}
\end{align} 
with $v_{\ell|k} = K z_{\ell|k}$ is satisfied for all $\ell = N + j$ with $j \in \N$. 
\end{theorem}

\begin{proof}
First, we notice that
\begin{equation}\label{eq:th1}
z_{N|k} \in \Ec_{W_x}\!(\alpha_N), \quad \text{with} \quad \alpha_N = \Big(z_{N|k}^\top W_x^{-1} z_{N|k}\Big)^{1/2}.
\end{equation}
Moreover, since $\Ec_{W_x}\!\left(\rho(1-\lambda^N)\right)$ is a probabilistic reachable set from \eqref{eq:reachLMI}, we have that
\begin{equation}
\Pr{e_{N|k} \in \Ec_{W_x}\!\left(\rho(1-\lambda^N)\right)} \geq 1-\EP(\rho),
\end{equation}
which implies that
\begin{equation}\label{eq:condN}
\Pr{x_{N|k} \in \Ec_{W_x}\!\left( \alpha_N + \rho (1-\lambda^N) \right)} \geq 1-\EP(\rho).
\end{equation}
Hence, similarly to Proposition~\ref{prop:xellk}, see \eqref{eq:prop1}, and relying on the definition of PRS, we have that \eqref{eq:constr_zNk} implies 
\begin{equation}\label{eq:th1b}
\Ec_{W_x}\!\left( \alpha_N + \rho (1-\lambda^N)\right) \subseteq \Ec_{W_x}\!\left( r_{N}\right) \subseteq \X_N,
\end{equation}
guaranteeing chance constraints satisfaction for $\ell=N$.

Then, from \cite[Proposition 4]{fiacchini2021probabilistic} and combining conditions \eqref{eq:Pre} and \eqref{eq:lambda}, for all $j \in \N$ we have
\begin{align}
&z_{N+j|k} \in \Ec_{W_x}\!(\alpha_N \lambda^j), \\
&\Pr{e_{N+j|k} \in \Ec_{W_x}\!\left( \rho(1-\lambda^{N+j})\right) } \geq 1-\EP(\rho), \label{eq:th1_a}
\end{align}
which implies that 
\begin{equation}\label{eq:condK1}
\Pr{x_{N+j|k} \in \Ec_{W_x}\!\left(\alpha_N \lambda^j + \rho(1-\lambda^{N+j}) \right)} \geq 1-\EP(\rho).
\end{equation}
It is now left to prove that \eqref{eq:condK1} implies \eqref{eq:constr_xk2} for all $\ell \geq N$. First, consider the case $\alpha_N < \lambda^N \rho$. Hence, we have
\begin{equation*}\label{eq:th1c1}
\Ec_{W_x}\!\left( \alpha_N \lambda^j + \rho(1-\lambda^{N+j})\right) \subseteq \Ec_{W_x}\!\left(\rho \right)
\end{equation*}
and then, from $\rho \leq r_{N}$ and \eqref{eq:condK1}, we get
\begin{equation*}\label{eq:th1c}
\Ec_{W_x}\!\left( \alpha_N \lambda^j + \rho(1-\lambda^{N+j})\right) \subseteq \Ec_{W_x}\!\left(\rho \right) \subseteq \Ec_{W_x}\!\left( r_{N}\right) \subseteq \XN,
\end{equation*}
implying constraints satisfaction with $v_{\ell|k} = K z_{\ell|k}$.
Now, considering the case 
\begin{equation}\label{eq:cond_decreas}
\alpha_N \geq \rho\lambda^{N},
\end{equation}
we search for a condition ensuring that \eqref{eq:condN} implies 
\begin{equation}\label{eq:condK}
\Pr{x_{N+j|k} \in \Ec_{W_x}\!\left(\alpha_N + \rho(1-\lambda^N) \right)} \geq 1-\EP(\rho), 
\end{equation}
for all $j \in \N$, see \eqref{eq:th1b}.
%
To guarantee the satisfaction of condition \eqref{eq:condK}, from \eqref{eq:condK1},  it is sufficient to have 
\begin{equation}
\alpha_N \lambda^j+ \rho(1-\lambda^{N+j}) \leq \alpha_N + \rho(1 - \lambda^{N}),
\end{equation}
for all $j \in \N$, condition holding if and only if \eqref{eq:cond_decreas} is satisfied.
Moreover, $\alpha_N$ satisfying \eqref{eq:constr_zNk} is equivalent to 
\begin{equation}\label{eq:cond_x11}
\alpha_N \leq r_{N} - \rho(1-\lambda^N).
\end{equation}
Therefore, combining \eqref{eq:cond_decreas} and \eqref{eq:cond_x11}, we obtain that $\Pr{x_{N+j|k} \in \XN} \geq 1-\EP(\rho)$ is satisfied for all $j \in \N$ if 
\begin{equation}
\lambda^{N} \rho \leq \alpha_N \leq r_{N} - \rho(1-\lambda^N), 
\end{equation}
which has an admissible solution if and only if \eqref{eq:constr_zNk} is satisfied with $\rho \leq r_{N}$, assumed holding. Consequently, to guarantee that both state and input chance constraints are satisfied by the system with control law $u_{\ell|k} = K x_{\ell|k}$ along the whole future trajectory starting at time $N$, it is sufficient to have \eqref{eq:constr_zNk} as terminal constraint for the nominal system, {from the definition \eqref{eq:XN} of $\XN$.}
\end{proof}

Clearly \eqref{eq:constr_xkXN}
implies that 
\begin{subequations}
\label{eq:constr_xk2}
\begin{align}
&\Pr{x_{\ell|k}\in \mathbb{X} | x_k} \geq 1-\EP(\rho),\label{eq:constr_xkx2}\\
&\Pr{u_{\ell|k}\in \mathbb{U} | x_k} \geq 1 - \EP(\rho),\label{eq:constr_xku2}
\end{align} 
\end{subequations}
with $v_{\ell|k} = K z_{\ell|k}$ are satisfied for all $\ell = N + j$ with $j \in \N$. 
Hence, similarly to the state and input tightened constraints, we can define a terminal region as follows
\begin{align}
    \label{eq:ZN}
\mathbb{Z}_N 
&=\Ec_{W_x}\!\big(r_N-\rho(1-\lambda^N)\big)
\end{align}
{which represents a positive invariant ellipsoid for the nominal dynamics.}

It is worth highlighting that condition \eqref{eq:constr_zNk} also ensures that $r_{N} - \rho(1-\lambda^N)$ is non-negative. Moreover, we have that condition \eqref{eq:cond_decreas} does not depend on $j$. Hence, if the set $z_{N|k} + \Ec_{W_x}\!( \rho(1-\lambda^N)),$ to which $x_{N|k}$ belongs with probability $1-\EP(\rho)$, is contained in a set $\text{big enough}$, then $z_{N+j|k} + \Ec_{W_x}\!(\rho(1-\lambda^{N+j}))$ will be included in the same set (or in smaller ones). This represents a reasonable condition for guaranteeing stability. Additionally, if the set 
$$\lim_{j \rightarrow +\infty} \Ec_{W_x}\!(\rho(1-\lambda^j)) = \Ec_{W_x}\!(\rho),$$
which is an outer approximation of the minimal probabilistic invariant set with violation probability $\EP(\rho)$, is not contained in $\mathbb{X}$, then the chance constraint on the state may be violated along the trajectory. {This gives a geometric meaning to constraint $\rho \leq r_N$.}

\begin{remark}
It is important to remark that the classical approach to guarantee recursive satisfaction of the terminal constraint would lead to a more conservative solution that the one proposed in this section. Indeed, a common choice in the standard approach consists in imposing that the final state shall belong to an invariant set for the nominal dynamics, obtained by subtracting a probabilistic invariant set
\begin{equation}\label{eq:PRS_ell_infty}
\mathcal{R}_\infty^x = \Ec_{W_x}\!\big(\rho\big)
\end{equation}
for instance, from the state constraint set $\XN$. This could be done by imposing 
\begin{equation}
z_{N|k} \in \Ec_{W_x}\!(r_N) \ominus \Ec_{W_x}\!(\rho) = \Ec_{W_x}\!(r_N - \rho),
\end{equation}
which is equivalent to 
\begin{equation}\label{eq:term_inv}
z_{N|k}^\top W_x^{-1} z_{N|k} \leq \left(r_N - \rho\right)^2.
\end{equation}
However, comparing \eqref{eq:constr_zNk} with \eqref{eq:term_inv}, it is evident the conservatism of the second (classical) approach, mainly for $\lambda$ close to one.\\
\end{remark}

\subsection{Terminal cost}\label{sec:term_cost}

The last element of the nominal optimization problem to be defined is the terminal cost $V_f(z_{N|k})$. Suppose that, without loss of generality\footnote{Note that such condition entails no loss of generality, since an appropriate value of $\nu$ always exists for it to hold.}, the weighting matrices $Q \in \S^n$ and $R \in \S^m$ in \eqref{eq:cost_infty} satisfy 
\begin{equation}\label{eq:WQR}
 Q + K^\top R K\preceq \nu W_x^{-1}.
\end{equation}
Consider the following function 
\begin{equation}\label{eq:Vf}
V_f(z) = \frac{{\nu}}{1-\lambda^2} \ z^\top W_x^{-1}z = \frac{{\nu}}{1-\lambda^2} \|z\|_{W_x^{-1}}^2
\end{equation}
as terminal cost for the SMPC problem \eqref{eq:SMPC_init} with $\nu~>~0$, 
being a Lyapunov function for the nominal system in closed loop with $v_{\ell|k} = K z_{\ell|k}$. In particular, from \eqref{eq:lambda}, for the nominal system \eqref{eq:sys_z}, we have 
\begin{equation}
\begin{aligned}
V_f\left(z_{\ell+1|k}\right) & - V_f(z_{\ell|k}) =  V_f\left((A + B K)z_{\ell|k}\right) - V_f(z_{\ell|k})\\
 & = \frac{\nu}{1-\lambda^2} \left(\|A_Kz_{\ell|k}\|_{W_x^{-1}}^2 - \|z_{\ell|k}\|_{W_x^{-1}}^2 \right) \\
& = -\frac{\nu}{1-\lambda^2}  \|z_{\ell|k}\|_{\left(W_x^{-1} - A_K^\top W_x^{-1} A_K\right)}^2\\
& \leq - \frac{\nu}{1-\lambda^2} \|z_{\ell|k}\|_{\left( 1 -\lambda^2 \right)W_x^{-1}}^2\\
& = -\nu \|z_{\ell|k}\|_{W_x^{-1}}^2 \leq - \|z_{\ell|k}\|_{\left( Q + K^\top R K \right)}^2,
\end{aligned}
\end{equation}
which is then exponentially decreasing and bounded above by the stage cost for $v_{\ell|k}=Kz_{\ell|k}$.

\section{SMPC formulations}
\label{sec:main_res}
The derivations of the previous section allow us to formulate an SMPC control scheme based on the definition of ellipsoidal-based PRS as tightened constraint sets which aim at guaranteeing recursive feasibility and chance constraints satisfaction. In this Section, we will show first how the proposed ellipsoidal PRS allows to guarantee \textit{initial-state conditioned RF}, as already done in other approaches (see e.g., \cite{hewing2018stochastic}), without the need to introduce any relaxation on the initial chance constraints. Then, we will present the main result of this paper, i.e., a two-phases SMPC based on ellipsoidal PRS and a relaxation backup strategy, which guarantees \textit{closed-loop} chance constraints satisfaction and \textit{measured-state conditioned RF in expectation}. The idea is rather classical: as a first step we consider an optimization problem that, if feasible, guarantees the required closed-loop chance constraints satisfaction. Whenever the problem is not feasible, we adopt a backup strategy which consists in properly relaxing the chance constraints so that the ensuing problem still guarantees the measured-state conditioned recursive feasibility in expectation as defined in Definition~\ref{def-MSRF}.

\subsection{An ellipsoidal-based approach guaranteeing ISRF}
In the framework of SMPC for system subject to chance constraints, we allow, by construction, a nonzero probability $\varepsilon$ of violating some of the constraints at (any) time $k$. Hence, feasibility at all steps $k$ cannot be guaranteed. A possible way out is to adopt the same philosophical approach proposed in \cite{farina2015approach,hewing2018stochastic}, and have initial-state conditioned recursive feasibility guarantees. That is, instead of initializing the nominal state $z_{0|k}$ with the (possibly unfeasible) measured state $x_k$ at time $k$, we adopt an open-loop initialization strategy with 
\begin{equation}
    \label{eq:BasicIND_x0}
z_{0|k} =  z_{1|k-1}.    
\end{equation}
In this setup, we can provide exactly the same theoretical guarantees given in \cite{farina2015approach,hewing2018stochastic} regarding initial-state conditioned recursive feasibility and closed-loop stability while employing the proposed ellipsoidal PRS to define the tightened constraints sets. This is formalized in the next Theorem, and the proof of this result is very similar to the one in \cite{hewing2018stochastic}, and it is reported in Appendix~\ref{app:proof3} for completeness.

\begin{theorem}\label{th:rec_feas}
Consider system  \eqref{eq:sys_ze} and constraint sets \eqref{eq:XU}.
Define the following problem
\begin{subequations}\label{eq:BasicSMPC}
\begin{align}
& \hspace{-0.6cm} \min_{\textbf{z}_k,\,\textbf{v}_k}\sum_{\ell = 0}^{N-1} \left(\|z_{\ell|k}\|_Q^2+ \|v_{\ell|k}\|_R^2\right)+V_f(z_{N|k})
\nonumber\\
\hspace{-0.2cm} \text{s.t.} \ & z_{\ell+1|k} = A z_{\ell|k} + B v_{\ell|k}, \\
&
z_{0|k}= z_{1|k-1},\label{eq:BasicSMPC_init}\\  
& z_{\ell|k}^\top W_x^{-1} z_{\ell|k} \leq (r_{x} - \rho(1-\lambda^\ell))^2, \;\; \ell\in\mathbb{N}_1^{N-1}, \label{eq:Basic_init_z}\\
& v_{\ell|k}^\top W_u^{-1} v_{\ell|k} \leq (r_{u} - \rho(1-\lambda^\ell))^2, \;\; \ell\in\mathbb{N}_1^{N-1},\\
& z_{N|k}^\top W_x^{-1} z_{N|k} \leq (r_{x} - \rho(1-\lambda^N))^2, 
\label{eq:Basic_end_zx}\\
& z_{N|k}^\top W_x^{-1} z_{N|k} \leq (r_{u} - \rho(1-\lambda^N))^2, \label{eq:Basic_end_zu}\\
& \rho \leq r_x,\quad \rho \leq r_u. 
\label{eq:Basic_end_r}
\end{align}
\end{subequations}   
If the optimization problem \eqref{eq:BasicSMPC} is feasible for a given initial condition $x_0 = z_{0|0}$, then it is recursively feasible.
Moreover, the resulting states $x_k$ and inputs $u_k$ satisfy the closed-loop chance constraints {\eqref{eq:constr_x0} and then Problem \eqref{eq:BasicSMPC} is initial-state conditioned recursive feasible.}
\end{theorem}

Hence, as a first intermediate result, we have shown how our proposed setup recovers the ISRF guarantees of \cite{hewing2018stochastic}. Similarly to the cited work, the approach allows to account for correlated noise. However, differently from \cite{hewing2018stochastic}, our formulation introduces in an explicit way the relaxation parameter $\rho$ that accounts for the desired probabilistic guarantees. More importantly, Problem \eqref{eq:BasicSMPC} introduces the two radii $r_x$ and $r_u$, which allow to ``control" the size of the respective tightened constraint sets. 
Notably, the dependence on these parameters is linear: this unique feature will be exploited in the next section where we show how such radii will play the role of \textit{relaxation} parameters.

Another possibility would be to rely on a ``dual-mode" approach in the spirit of the two-modes design proposed in \cite{hewing2018stochastic}.  In particular, one can substitute  {\eqref{eq:BasicSMPC_init}} with the following case-dependent initialization
\begin{equation}\label{eq:2modes_x0}
z_{0|k}= 
\left\lbrace\begin{array}{ll}
x_k &\text{if \eqref{eq:BasicSMPC} is feasible with }  {z_{0|k} = x_k } \\
 z_{1|k-1} & \text{otherwise.}
\end{array}   \right. 
\end{equation}
In other words, this mixed-approach can be interpreted in this way: whenever the current state is not feasible, we disregard the unfeasibility and resort on the guarantees we had at the previous feasible step relying on an open-loop strategy. Of course, as already commented, this approach may provide a false sense of security, and it may be considered in some sense sub-optimal, since it practically disregards the information provided by the current measurement. In turn, as observed  in \cite{paulson2020stochastic}, this ``may degrade closed-loop performance when the states are not in the region of attraction of the controller".

As an alternative, in the next section we show how it is possible to guarantee closed-loop chance constraints satisfaction and to recover recursive feasibility (specifically MSRF) without sacrificing the information carried by the state measurement at current time, eventually relying on a backup strategy based on the relaxation of radius $r_x$ and $r_u$ .

\subsection{Ellipsoidal-based approach guaranteeing MSRF}
Let us assume that a value of $\rho>0$ is given, i.e., a desired minimal probability bound for the chance constraints to be satisfied. Then, we define the following SMPC problem.

\begin{definition}[Ellipsoidal Tube SMPC]
Given the system  \eqref{eq:sys_ze} and constraint sets \eqref{eq:XU},  select
$W_x\succ 0$ and $\lambda \in (0,1)$ satisfying \eqref{eq:reachLMI}. Then, compute $r_x$ and $(r_u,W_u)$ according to \eqref{eq:rx}, \eqref{eq:Wu}, respectively, and solve the following finite horizon conic programming problem 
\begin{equation}
\begin{aligned}
\min_{\textbf{z}_k, \textbf{v}_k} &\sum_{\ell = 0}^{N-1} \left(\|z_{\ell|k}\|_Q^2+ \|v_{\ell|k}\|_R^2\right)+V_f(z_{N|k})
\label{eq:BasicSMPC_x0}
\\
\text{s.t.} \ \ & z_{\ell+1|k} = A z_{\ell|k} + B v_{\ell|k}, \\
&
z_{0|k}= 
 x_{k}\\  
& 
\eqref{eq:Basic_init_z}-\eqref{eq:Basic_end_r}. 
\end{aligned}
\end{equation}
\end{definition}

The above problem, \textit{whenever it is feasible}, ensures that
\begin{align*}
&\Pr{x_{\ell|k}\in \mathbb{X} | \ x_k} \geq 1-\varepsilon_x,\\
&\Pr{u_{\ell|k}\in \mathbb{U} | \ x_k} \geq 1 - \varepsilon_u,
\end{align*} 
hold for all $\ell \in \N$, that is \eqref{eq:constr_xk}
holds without need of relaxing the sets. However, it is clear that we cannot guarantee that problem  {\eqref{eq:BasicSMPC_x0}} will be feasible at all steps $k$, since we allow, by construction, a nonzero probability $\varepsilon$ of violating some of the constraints at time $k$. 

In case  {Problem~\eqref{eq:BasicSMPC_x0}} is unfeasible for the current measure $x_k$, then we define a backup strategy based on the relaxation of the tightening bounds defined by $r_x$ and $r_u$ to admit a solution. And typically, the introduced relaxation is expected to be the minimum one. Thus, if  {Problem~\eqref{eq:BasicSMPC_x0}} is infeasible at some time instant $k$, it is possible to solve the following receding horizon problem as backup scheme including an ad-hoc relaxation over $r_{x}$ and $r_{u}$ and guaranteeing feasibility also for the relaxed problem
\begin{subequations}\label{eq:SMPC_barr}
\begin{align}
\hspace{-0.2cm} \Delta r^\star & (x_k) = \min_{\textbf{z}_k, \textbf{v}_k, \bar{r}_x, 
\bar{r}_u} \max \{\bar{r}_x - r_x, \ \bar{r}_u - r_u, \ 0 \} \label{eq:cost-barr}\\
\hspace{-0.2cm} \text{s.t.} \ \ & z_{\ell+1|k} = A z_{\ell|k} + B v_{\ell|k}, \label{eq:con1bar}\\ 
& z_{0|k}=x_k\\ 
& z_{\ell|k}^\top W_x^{-1} z_{\ell|k} \leq (\bar{r}_x - \rho(1-\lambda^\ell))^2, \; \ell\in\mathbb{N}_1^{N-1}, \label{eq:prabar}\\
& v_{\ell|k}^\top W_u^{-1} v_{\ell|k} \leq (\bar{r}_u - \rho(1-\lambda^\ell))^2, \; \ell\in\mathbb{N}_1^{N-1}, \label{eq:prbbar}\\
& z_{N|k}^\top W_x^{-1} z_{N|k} \leq (\bar{r}_x - \rho(1-\lambda^N))^2, \label{eq:prcbar}\\
& z_{N|k}^\top W_x^{-1} z_{N|k} \leq (\bar{r}_u - \rho(1-\lambda^N))^2, \label{eq:prcbar2}\\
& \rho \leq r_x \leq \bar{r}_x, \label{eq:rxbar}\\
& \rho \leq r_u \leq \bar{r}_u, \label{eq:rubar}
\end{align}
\end{subequations}

The following theorem proves that Problem \eqref{eq:SMPC_barr} is always feasible and the MSRF in expectation is guaranteed.

\begin{theorem}\label{th:feasibility}
Denote with $\textbf{z}^\star(x_k)$, $\textbf{v}^\star(x_k)$, $\bar{r}_x^\star(x_k)$, and $\bar{r}_u^\star(x_k)$ the optimal solution of problem \eqref{eq:SMPC_barr} and $u_k^\star(x_k) = \textbf{v}_{0|k}^\star(x_k)$ the control input applied to the system \eqref{eq:sys} at time $k$. Suppose that $\rho$ is such that 
\begin{equation}\label{eq:th1boundrho}
\rho\geq \sqrt{\frac{n(1-\lambda)}{1+\lambda}},
\end{equation}
then the following properties hold:
\begin{itemize}
\item[(i)] Problem \eqref{eq:SMPC_barr} is feasible for all $x_k \in \R^n$ at time $k \in \N$. 
\item[(ii)] The solution of Problem~\eqref{eq:SMPC_barr} for $x_{k+1}$ at time $k+1$ is such that 
\begin{align}
& \E{\bar{r}_u^\star(x_{k+1}) \, | \, x_k} \leq \bar{r}_u^\star(x_{k})\label{eq:condru}\\
& \E{\bar{r}_x^\star(x_{k+1}) \, | \, x_k} \leq \bar{r}_x^\star(x_{k}) \label{eq:condrx}\\
& \E{\Delta r^\star(x_{k+1}) \, | \, x_k} \leq \Delta r^\star(x_{k}) \label{eq:condr}
\end{align}
\item[(iii)] Problem \eqref{eq:SMPC_barr} is measured-state conditioned recursively feasible in expectation,
{and \eqref{eq:RCFE_def}-\eqref{eq:constr_xk} hold with 
\begin{align}
\gamma_u^\star(x)&=\bar r_u^\star(x)/r_u,  \qquad
\gamma_x^\star(x) = \bar r_x^\star(x)/r_x. \label{eq:gammar}
\end{align}
}
\end{itemize}
\end{theorem}

Proof of Theorem \ref{th:feasibility} is reported in Appendix~\ref{app:proof6}.

\begin{remark}
{Note that condition \eqref{eq:th1boundrho} is not restrictive. Indeed, in the case of non Gaussian noise, the condition $n/\rho^2\le\varepsilon$ arising from \eqref{eq:eps_r} automatically implies satisfaction of~\eqref{eq:th1boundrho}, since
\[
\rho\ge\sqrt{\frac{n}{\varepsilon}}\ge\sqrt{n}\ge\sqrt{\frac{n(1-\lambda)}{1+\lambda}}.
\]
}
\end{remark}

\subsection{Measured-state dependent SMPC}
If one want to solve a single optimization problem, one possibility is to rely on a novel SMPC formulation that properly balances recursive probabilistic guarantees and feasibility. In particular, we introduce the following definition of  {measured-state dependent SMPC (MS-SMPC)}.

\begin{definition}[ {MS-SMPC}]
Given the system  \eqref{eq:sys_ze} and constraint sets \eqref{eq:XU}, select
$W_x\succ 0$ and $\lambda \in (0,1)$ satisfying \eqref{eq:reachLMI}. Then, compute $r_x$   and $(r_u,W_u)$ according to \eqref{eq:rx}, \eqref{eq:Wu}, respectively, and solve the following finite horizon optimization problem 
\begin{subequations}
\label{eq:SMPC_mixed}
\begin{align}
&\hspace{-0.3cm}  \min_{\textbf{z}_k, \textbf{v}_k, \bar{r}_x, 
\bar{r}_u} 
\sum_{\ell = 0}^{N-1} \left(\|z_{\ell|k}\|_Q^2+ \|v_{\ell|k}\|_R^2\right) +V_f(z_{N|k}) \nonumber\\ 
& \hspace{2.3cm} +\,\,\mu \max\{\bar{r}_x - r_x, \ \bar{r}_u - r_u, \ 0 \} \label{eq:cost-balanced}\\
&\hspace{-0.3cm} \text{s.t.} \ z_{\ell+1|k} = A z_{\ell|k} + B v_{\ell|k},\\ 
&\hspace{-0.3cm} \phantom{s.t.} \ z_{0|k} = x_k,\\ 
&\hspace{-0.3cm} \phantom{s.t.} \ z_{\ell|k}^\top W_x^{-1} z_{\ell|k} \leq (\bar{r}_x - \rho(1-\lambda^\ell))^2, \; \ell\in\mathbb{N}_1^{N-1}, \label{eq:prabarB}\\
&\hspace{-0.3cm} \phantom{s.t.} \ v_{\ell|k}^\top W_u^{-1} v_{\ell|k} \leq (\bar{r}_u - \rho(1-\lambda^\ell))^2, \; \ell\in\mathbb{N}_1^{N-1}, \label{eq:prbbarB}\\
&\hspace{-0.3cm} \phantom{s.t.} \ z_{N|k}^\top W_x^{-1} z_{N|k} \leq (\bar{r}_x - \rho(1-\lambda^N))^2, \label{eq:prcbarB}\\
&\hspace{-0.3cm} \phantom{s.t.} \ z_{N|k}^\top W_x^{-1} z_{N|k} \leq (\bar{r}_u - \rho(1-\lambda^N))^2, \label{eq:prcbar2B}\\
&\hspace{-0.3cm} \phantom{s.t.} \ \rho \leq r_x \leq \bar{r}_x, \label{eq:rxbarB}\\
&\hspace{-0.3cm} \phantom{s.t.} \ \rho \leq r_u \leq \bar{r}_u, \label{eq:rubarB}
\end{align}
\end{subequations}
\end{definition}

Note that, although the cost is not differentiable, the problem \eqref{eq:SMPC_mixed} is still convex. Moreover, it is possible to consider smooth approximations, e.g., by replacing the $\max$ in \eqref{eq:SMPC_mixed} with $\left(\bar{r}_x - r_x\right)^{\beta} + \left(\bar{r}_u - r_u\right)^{\beta}$, with $\beta$ big enough.

The rationale behind Problem~\eqref{eq:SMPC_mixed} is simple: as long as Problem \eqref{eq:BasicSMPC_x0} is feasible, conditions \eqref{eq:prabarB}-\eqref{eq:rubarB} hold for $\bar{r}_x=r_x$ and $\bar{r}_u=r_u$, and then \eqref{eq:cost-balanced} boils down to the cost of \eqref{eq:BasicSMPC_x0}, in which case, Problem \eqref{eq:SMPC_mixed} is equivalent to Problem \eqref{eq:BasicSMPC_x0}. On the other end, whenever the current value of $x_k$ is such that Problem \eqref{eq:BasicSMPC_x0} is not feasible, with $\mu$ big enough, the second term in \eqref{eq:cost-balanced} would dominate the first one, leading to the feasible solution with minimal mismatch between $\bar{r}_x$ and $r_x$, and between $\bar{r}_u$ and $r_u$, then minimizing the constraints relaxation and recovering Problem~\eqref{eq:SMPC_barr}.

\begin{remark}
The idea of guaranteeing feasibility by relaxing the constraints has also been used in \cite{paulson2020stochastic}, for treating the problem of hard input constraints in SMPC. However, it shall be noted that, in the case of \cite{paulson2020stochastic}, the resulting problem was not \textit{jointly convex} in the design and relaxation parameters, and an iterative two-stage optimisation strategy needed to be adopted to tackle the problem. In the setup we propose, both the constraints scaling parameters $\bar{r}_x$ and $\bar{r}_u$, but also the probability bound $\rho$, appear linearly in the convex optimization problem. This ensures the optimality of the solution since, by adopting an exact penalty function method, the backup controller yields the same solution as the fully constrained MPC problem  {\eqref{eq:BasicSMPC_x0}}, when the latter is feasible.
\end{remark}

Besides the ensured feasibility of Problem~\eqref{eq:SMPC_mixed}, we can establish an asymptotic average performance bound for Problem \eqref{eq:SMPC_mixed}, based on a cost decrease in expectation.

\begin{theorem}\label{th:balanced_cost}
Denoting $J_{r}^\star(x_k)$ the optimal value function of \eqref{eq:SMPC_mixed}, then 
    \begin{equation}
\mathbb{E}\{J_{r}^\star(x_{k+1})\,|\,x_k\}-J_{r}^\star(x_k) \leq -\|x_{k}\|_Q^2-\|u_{0|k}\|_R^2,
    \end{equation}
holds for every $x_k$.
\end{theorem}

Proof of  Theorem~\ref{th:balanced_cost} is reported in Appendix~\ref{app:proof5}.

\section{State dependent probability bounds}
\label{sec:pr_state}
Given the solutions $\mathbf{z}^\star_{k},\,\mathbf{v}^\star_{k}$, $\bar{r}_x^\star$, and $\bar{r}_u^\star$ of Problem \eqref{eq:SMPC_mixed}, we can provide different guaranteed bounds on the probability of constraints satisfaction, under specific conditions.

\begin{remark}
In \eqref{eq:SMPC_mixed}, the constraint bounds are softened introducing $\bar{r}_{x}$ and $\bar{r}_{u}$ as optimization variables. Analogously, one could envision to consider parameter $\rho$ as the free optimization variable while maintaining $\bar{r}_{x} = r_{x}$ and $\bar{r}_{u} = r_{u}$, hence allowing relaxations of the violation probability bound rather than the constraints relaxation. Note that $\bar{r}_x$, $\bar{r}_u$, and $\rho$ appear linearly in Problem~\eqref{eq:SMPC_mixed}. Thus, all of them could have been considered as optimization parameters without any relevant complexity increase. Hereafter, though, we only use the relation between the parameters to have tighter \textnormal{a posteriori} estimations of the violation bound $\rho$.
\end{remark}

First, let us consider the state constraints. For every element of the optimal sequence  $z^\star_{\ell|k}$ two possibilities exist, either $z_{\ell|k}^\top W_x^{-1} z_{\ell|k} < r_x^2$ or not. 

In the first case, i.e., $z_{\ell|k}^\top W_x^{-1} z_{\ell|k} < r_x^2$, the nominal state satisfies $z_{\ell|k} \in \Ec_{W_x}\!(r_{x})$ and consequently the probability for $x_{\ell|k} \in \Ec_{W_x}\!(r_{x})$ can be computed by using the Chebychev bound. Indeed, a value $\bar{\rho}_{\ell|x} > 0$ exists such that 
\begin{align}\label{eq:bound_probability}
\sqrt{z_{\ell|k}^\top W_x^{-1} z_{\ell|k}} = r_{x} - \bar{\rho}_{\ell|x}(1-\lambda^\ell), 
\end{align}
which implies that $x_{\ell|k} \in \Ec_{W_x}\!(r_{x}) \subseteq \mathbb{X}$ with probability $1-\EP(\bar{\rho}_{\ell|x})$. Therefore, $1 - \EP(\bar{\rho}_{\ell|x})$ with 
\begin{equation}\label{eq:barrxk}
\bar{\rho}_{\ell|x} = \frac{r_{x} - \sqrt{z_{\ell|k}^\top W_x^{-1} z_{\ell|k}}}{1-\lambda^\ell},
\end{equation}
is the maximum lower bound on the probability for $x_{\ell|k}~\in~\Ec_{W_x}\!(r_{x})$, given the specific solution of the optimal problem. 

In the second case, i.e., if $z_{\ell|k}^\top W_x^{-1} z_{\ell|k} \geq r_x^2$, we have that $z_{\ell|k}\not\in\Ec_{W_x}\!(r_{x})$ or it is in its boundary. Therefore, the Chebychev bound cannot be applied to compute a guaranteed probability of constraints satisfaction. Indeed, in this case no positive values of $\bar{\rho}_{\ell|x}$ exists such that \eqref{eq:bound_probability} holds since neither the nominal value $z_{\ell|k}$ strictly satisfies the constraints.

Analogous considerations hold for the input constraints: if $v_{\ell|k}^\top W_u^{-1} v_{\ell|k} < r_u^2$ is satisfied, then a guaranteed lower bound on the probability for $u_{\ell|k}~\in~\Ec_{W_u}(r_{u}) \subseteq \U$ is given by $1 - \EP(\bar{\rho}_{\ell|u})$ with 
\begin{equation}\label{eq:barruk}
\bar{\rho}_{\ell|u} = \frac{r_{u} - \sqrt{v_{\ell|k}^\top W_u^{-1} v_{\ell|k}}}{1-\lambda^\ell}.
\end{equation}

Finally, if $z_{N|k}^\top W_x^{-1} z_{N|k} < r_N^2 = (\min \{r_x,r_u\})^2$, then $x_{N|k} \in \Ec_{W_x}(r_N)$ with  probability at least $1 - \EP(\bar{\rho}_{N})$, where
\begin{equation}\label{eq:barrxN}
\bar{\rho}_{N} = \frac{r_N - \sqrt{z_{N|k}^\top W_x^{-1} z_{N|k}}}{1-\lambda^N}.
\end{equation}

\section{Initial input bounds}\label{sec:inputbounds}
Concerning the bounds on the initial input $v_{0|k}$, we outline three possible choices, each one providing a specific feature to the optimization problem, as detailed in the follows: a) no bounds; b) hard bounds; and c) soft bounds. 

\subsection{Case A - no bounds}\label{ssect:A}
In case \textit{no bounds} are enforced on $v_{0|k}$, this strategy would lead to more aggressive initial inputs that will drive the nominal states along the prediction to remain inside the region where chance constraints are guaranteed. As a consequence, this approach would imply larger input violation at the first step and lower violation occurrences along the (predicted) trajectory. 

\subsection{Case B - hard bounds}\label{ssect:B}
The second possibility is to enforce \textit{hard bounds} on the first input such that 
$v_{0|k} \in \mathbb{U}$, by simply imposing $H_u v_{0|k} \leq h_u$. This choice would automatically exclude any input constraint violation at the first step, being $v_{0|k}= u_k$ the input to be applied in the MPC strategy control. On the other hand, we could still have input violations along the nominal predicted trajectory for $\ell\in\mathbb{N}_1^{N-1}$. Moreover, this would imply that the recursive feasibility and the estimated tightening bounds are no more valid, since the input violation, admitted within the prediction, is not allowed for the MPC application.  {In other words, the problem solved at time $k+1$ has hard constraints on the first nominal input $v_{0|k+1}$ that were not imposed on $v_{1|k}$.} Hence, the prediction obtained at time $k$ might not be admissible at time $k+1$. This mismatch between prediction and realization appears evidently in the predicted and realized trajectories, depicted in Figure \ref{fig:caseA} and Figure \ref{fig:alltraj}, respectively.

\subsection{Case C - soft bounds}\label{ssect:C}
The third case, that can be seen as a compromise between the two previous approaches, consists in using the relaxation variable $\gamma_{u}(x_k)$ to \textit{soften} also the bounds on the initial input, i.e., $v_{0|k} \in \gamma_{u}(x_k)\mathbb{U}$. This method implies the possibility of relaxation of the initial input bound, provided that there is no feasible solution for the original constraint, i.e., $\gamma_{u}(x_k)=1$, and the \textit{soft bound} can be obtained by adding the following constraint
\begin{equation}\label{eq:softconstr}
H_u v_{0|k} \leq \gamma_u(x_k)h_u = \bar{r}_{u}/r_{u} \ h_u,
\end{equation}
to Problem \eqref{eq:SMPC_mixed}.

\section{Numerical simulations}
\label{sec:num_sim}
To illustrate our approach, we consider a simple double integrator system, also used for example in \cite{hewing2018stochastic,hewing2020recursively,arcari2023stochastic} with 
\begin{equation}
A = \begin{bmatrix}
1 && 1\\
0 && 1
\end{bmatrix}, \quad 
B = \begin{bmatrix}
0.5\\
1
\end{bmatrix}.
\end{equation}
The local feedback gain $K = \left[0.2068 \ \, 0.6756\right]$ is the infinite horizon (discrete) LQR solution with $Q = \mathbb{I}_2$ and $R = 10$. The covariance matrix $\Gamma_w$ of the i.i.d. Gaussian process $w_k$, with null mean, and the matrix $W_x$, which satisfies the conditions \eqref{eq:reachLMI_1} and \eqref{eq:reachLMI_2} with $\lambda = 0.7503$, are given by 
\begin{equation*}
\Gamma_w = \left[\begin{array}{cc}
0.1 & 0.05\\
0.05 & 0.1
\end{array}\right], \quad W_x = \left[\begin{array}{cc}
10.9264  & -3.7386\\
-3.7386  &  3.8143
\end{array}\right].
\end{equation*}
The polytopic state and input constraint sets are
\begin{align*}
& \X = \{x \in \R^2: H_x x \leq h_x\} = \{x \in \R^2: \|x\|_\infty \leq 40\}\\
& \U = \{u \in \R: H_u u \leq h_u\} = \{u \in \R: |u| \leq 10\}
\end{align*}
giving $r_{x} = 12.1010$ as the maximal value such that $\Ec_{W_x}\!(r_{x}) \subseteq \X$, and $W_u = 0.2237$ with $r_{u} = 21.1448$ the maximal value such that $\Ec_{W_u}\!(r_u) \subseteq \U$ and \eqref{eq:WuWx} is satisfied. The selected violation probability level is $\varepsilon~=0.1$, leading to $\rho = 2.146$ that satisfies \eqref{eq:th1boundrho}. The minimum value of $\nu$ such that \eqref{eq:WQR} is satisfied, i.e., $\nu = 16.0082$, is used for the terminal cost \eqref{eq:Vf}. 

{For comparison purposes, we contrast our approach with a standard dual-mode open-loop strategy, in the spirit of \cite{farina2013probabilistic,hewing2018stochastic}. In particular, we consider the problem \eqref{eq:SMPC_init} with constraints given by \eqref{eq:ominus}, probabilistic sets \eqref{eq:PRS_ell}, \eqref{eq:PRS_ell_u}, and \eqref{eq:PRS_ell_infty}, and a dual-mode initialization approach, analogous to \eqref{eq:2modes_x0}. Note that for this problem, referred to as initial-state SMPC (IS-SMPC) in the follows, the ISRF conditions is ensured if it is feasible at time zero.}

\subsection{Comparison of input bounds strategies}

In our first set of simulations, to better highlight the effect of the proposed relaxation of {MS-SMPC} with respect to {IS-SMPC}, we select  as initial state the point $x_0 = (-40, \, 40)$, which is on the boundary of the constraint set $\X$. In this case, we have that this selected initial condition leads to infeasibility of {IS-SMPC}. Hence, the problem cannot be tackled by classical open-loop strategies. On the other hand, the proposed {MS-SMPC} is able to overcome the feasibility issue through the relaxation of the constraints by employing values of $\bar{r}_{x}$ and $\bar{r}_{u}$, greater than $r_{x}$ and $r_{u}$, respectively, in \eqref{eq:SMPC_mixed}. 

\begin{figure}[!ht]
\centering
\subfigure[case A]{\includegraphics[width=.8\columnwidth]{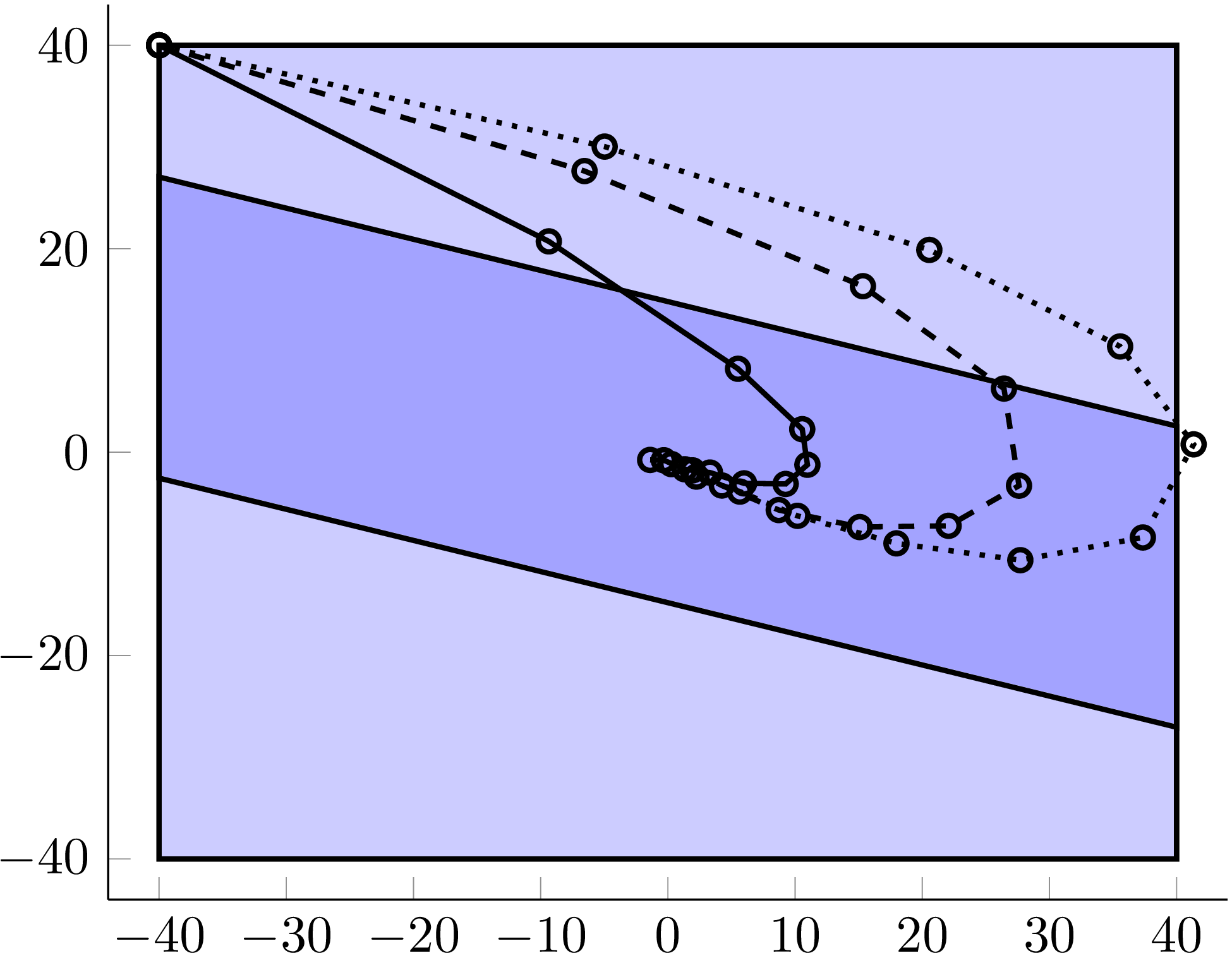}
\label{fig:caseA0}
}
\hfil
\subfigure[case B]{\includegraphics[width=.8\columnwidth]{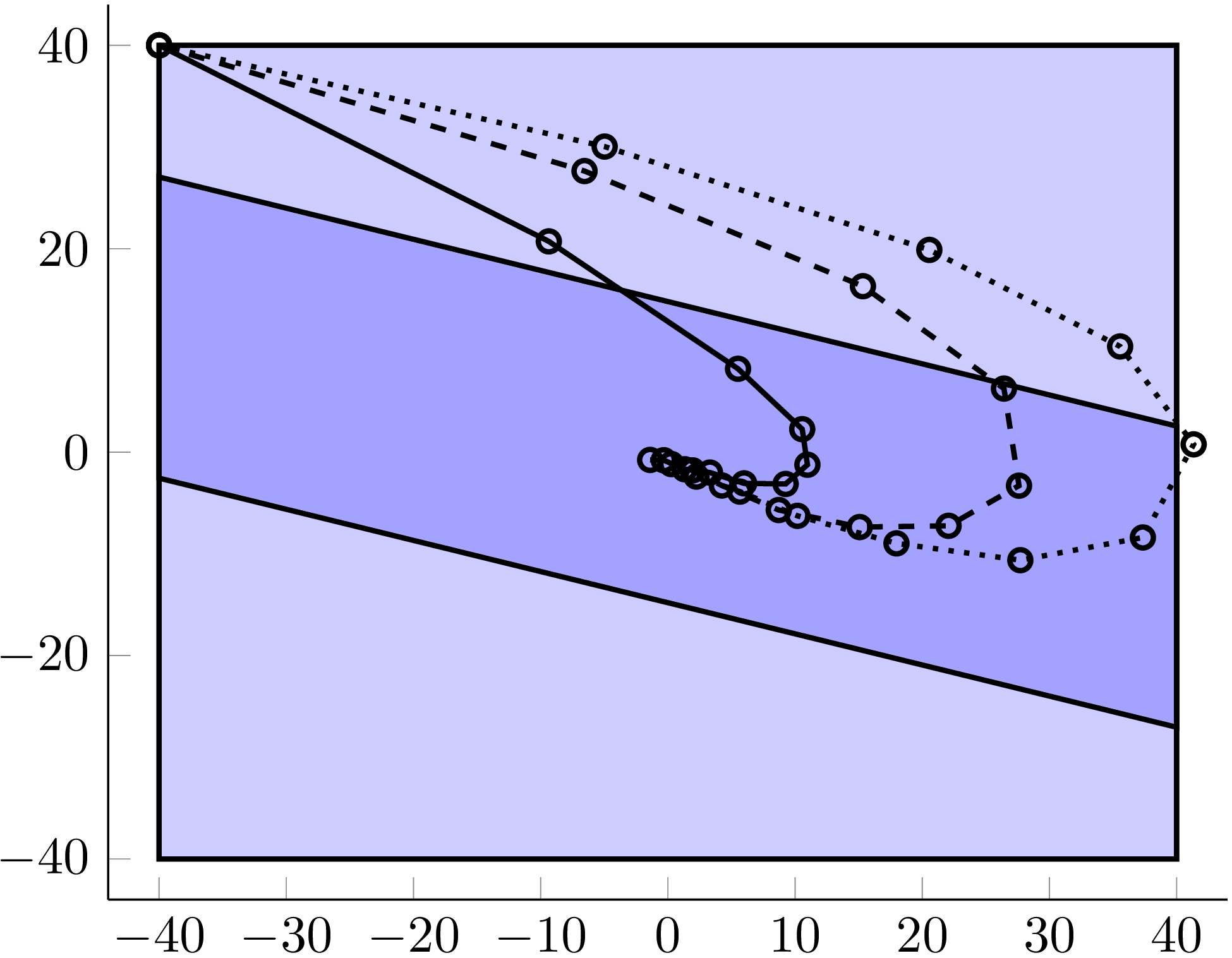}	\label{fig:caseB0}
}
\subfigure[case C]{\includegraphics[width=.8\columnwidth]{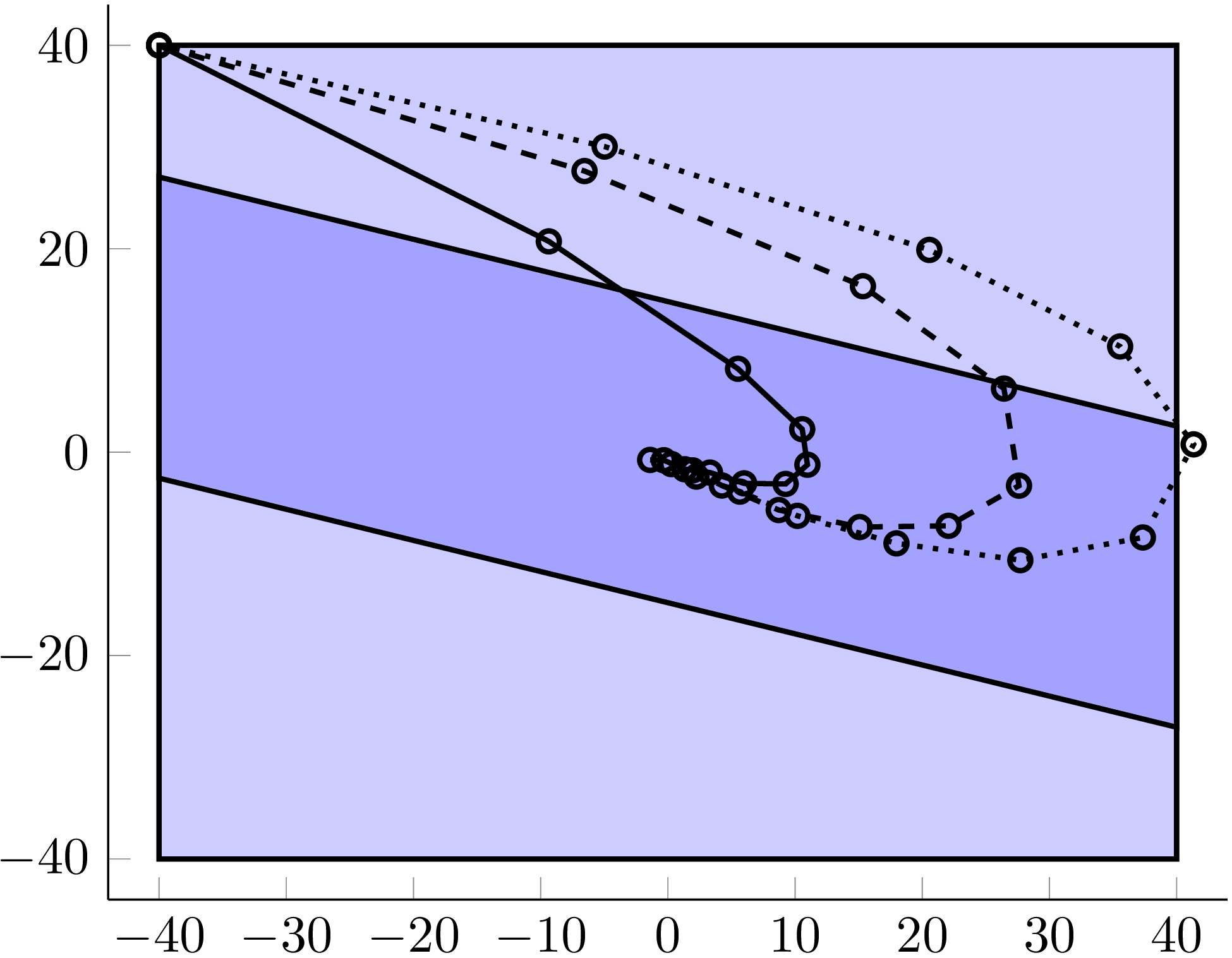}	\label{fig:caseC0}
}
\caption{Predicted state trajectories and sets $\Ec_{W_x}\!(\bar{r}_x - \rho(1-\lambda^\ell))$, starting from $x_0=(-40,\,40)$ with $\ell\in\mathbb{N}_0^{10}$, enforcing on the first input $v_0$: no bounds (top), hard bounds (middle), and soft bounds (bottom).}
\label{fig:caseA}
\end{figure}

Now, we proceed analysing the effects of the different assumptions on the initial input strategies described in Section \ref{sec:inputbounds}, when applied to the aforementioned system, namely by enforcing no bounds (case A), hard bounds (case B), or soft bounds (case C) on $v_{0|k}$ and solving the related optimization problem \eqref{eq:SMPC_mixed} with $N = 10$. 

Figure \ref{fig:caseA} depicts the predicted trajectories given by $z_{\ell|k}$ and the sets $\Ec_{W_x}\!(\bar{r}_x - \rho(1-\lambda^\ell))$ for $\ell \in \N_1^{N}$ for the three cases.
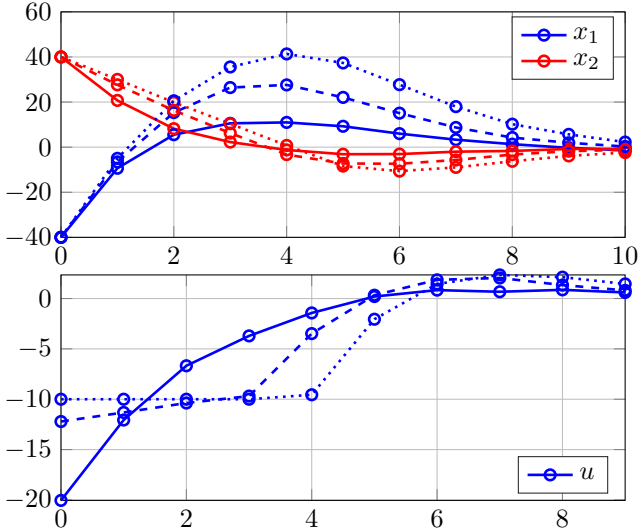
\begin{figure}[!ht]
\begin{center}
%
%
\begin{tikzpicture}

\begin{axis}[%
width=7.5cm,
height=3cm,
at={(0.0cm,3.5cm)},
scale only axis,
xmin=0,
xmax=10,
ymin=-40,
ymax=60,
axis background/.style={fill=white},
xmajorgrids,
ymajorgrids,
legend style={legend cell align=left, align=left, draw=white!15!black}
]
\addplot [color=blue, solid, line width=1pt, mark=o, mark options={solid, blue}]
  table[row sep=crcr]{%
0	-40\\
1	-9.3585028142666\\
2	5.51767264717709\\
3	10.5629937718162\\
4	10.9719102963344\\
5	9.25892363096628\\
6	5.99418388727221\\
7	3.30928824847435\\
8	1.32177595088701\\
9	-0.31813900936373\\
10	-1.39327410751767\\
};
\addlegendentry{$x_1$}

\addplot [color=red, dotted, line width=1pt, mark=o, mark options={solid, red}, forget plot]
  table[row sep=crcr]{%
0	40\\
1	30.0426357376669\\
2	19.8832463980677\\
3	10.3879896106416\\
4	0.764762161600457\\
5	-8.39145679214616\\
6	-10.6235760097651\\
7	-8.97313357609425\\
8	-6.26312204617795\\
9	-3.88253076117011\\
10	-2.4040733744948\\
};

\addplot [color=red, solid, line width=1pt, mark=o, mark options={solid, red}]
  table[row sep=crcr]{%
0	40\\
1	20.7203354740707\\
2	8.19372389284068\\
3	2.25637729003613\\
4	-1.24026929274027\\
5	-3.12942432278591\\
6	-3.07245023961244\\
7	-2.00934174388728\\
8	-1.68212278471176\\
9	-0.795751328609192\\
10	-0.767041628031613\\
};
\addlegendentry{$x_2$}

\addplot [color=blue, dotted, line width=1pt, mark=o, mark options={solid, blue}, forget plot]
  table[row sep=crcr]{%
0	-40\\
1	-4.96894666634425\\
2	20.5555337445424\\
3	35.5571229497868\\
4	41.3369210346659\\
5	37.3450705373759\\
6	27.7108800449181\\
7	17.9726491391822\\
8	10.1890582850044\\
9	5.66159763132641\\
10	2.24596701224095\\
};

\addplot [color=blue, dashed, line width=1pt, mark=o, mark options={solid, blue}, forget plot]
  table[row sep=crcr]{%
0	-40\\
1	-6.55111211004422\\
2	15.3349048804532\\
3	26.423701740268\\
4	27.6064577784954\\
5	22.069307460015\\
6	15.0810012586531\\
7	8.71953394363248\\
8	4.24482587905148\\
9	1.93960093074201\\
10	0.276478708164053\\
};
\addplot [color=red, dashed, line width=1pt, mark=o, mark options={solid, red}, forget plot]
  table[row sep=crcr]{%
0	40\\
1	27.6363812407324\\
2	16.3144135829931\\
3	6.24784837752746\\
4	-3.2970349043112\\
5	-7.24388654598225\\
6	-7.36621576295183\\
7	-5.69299053890974\\
8	-3.28084749903115\\
9	-1.79103315183045\\
10	-1.17617601786355\\
};

\end{axis}

\begin{axis}[%
width=7.5cm,
height=3cm,
at={(0cm,0cm)},
scale only axis,
xmin=0,
xmax=9,
ymin=-20.0477129823669,
ymax=2.34512869770875,
axis background/.style={fill=white},
xmajorgrids,
ymajorgrids,
legend pos=south east,
legend style={legend cell align=left, align=left, draw=white!15!black}
]
\addplot [color=blue, solid, line width=1pt, mark=o, mark options={solid, blue}]
  table[row sep=crcr]{%
0	-20.0477129823669\\
1	-12.0641159282694\\
2	-6.67753483301098\\
3	-3.70954449379207\\
4	-1.43156649763836\\
5	0.199168145187533\\
6	0.836018053926002\\
7	0.673098070317212\\
8	0.863134353556482\\
9	0.60346714238145\\
};
\addlegendentry{$u$}

\addplot [color=blue, dotted, line width=1pt, mark=o, mark options={solid, blue}, forget plot]
  table[row sep=crcr]{%
0	-9.99999999954731\\
1	-9.99999999374598\\
2	-9.9999999512041\\
3	-9.99999999723729\\
4	-9.55467025162155\\
5	-2.0552030366398\\
6	1.44588850619437\\
7	2.34512869770875\\
8	2.12411845938333\\
9	1.45216139940581\\
};
\addplot [color=blue, dashed, line width=1pt, mark=o, mark options={solid, blue}, forget plot]
  table[row sep=crcr]{%
0	-12.2131900051978\\
1	-11.3153332070849\\
2	-10.3756259291654\\
3	-9.68744986840332\\
4	-3.48283826577566\\
5	0.329276741786579\\
6	1.85757971922099\\
7	2.04289618158387\\
8	1.33869984507013\\
9	0.808923165069144\\
};
\end{axis}
\end{tikzpicture}%
\end{center}
\caption{State and input trajectories for SMPC enforcing no bounds (solid lines), hard bounds (dotted lines), and soft bounds (dashed lines) on the first input $v_{0|k}$.}
\label{fig:allstatein} 
\end{figure}
In case of no constraint, the {MS-SMPC} generates a predicted trajectory that reaches quickly the set $\X_K$ where both state and control input can be satisfied (see Figure~\ref{fig:caseA0}), and the value $\bar{r}_{x}$ and $\bar{r}_{u}$ are close to $r_{x}$ and $r_{u}$, respectively.
\begin{figure}[!ht]
\begin{center}
\includegraphics[width=0.8\columnwidth]{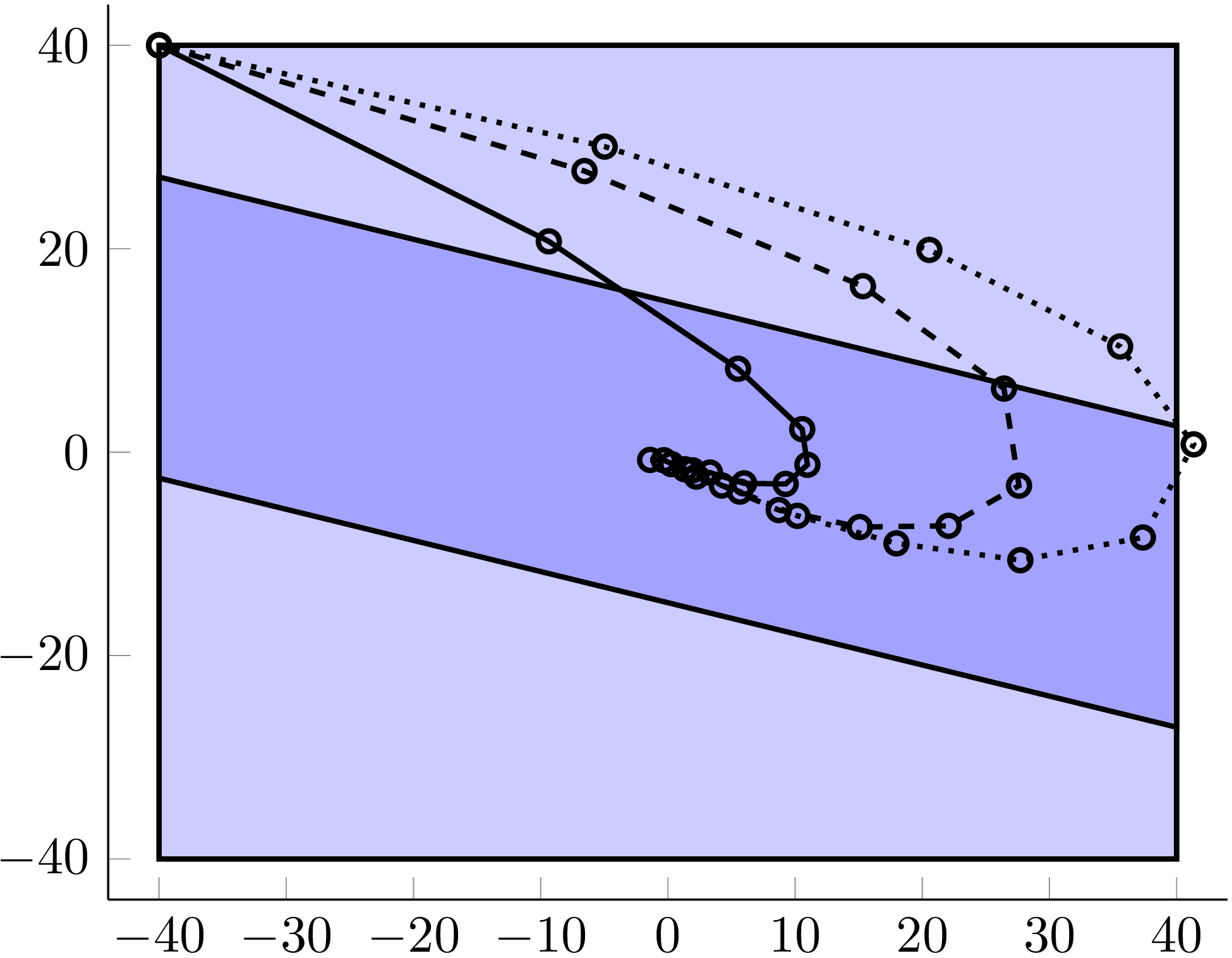}
\end{center}
\caption{Comparison among the realized state trajectories starting from $x_0=(-40,\,40)$ for $k \in [0,10]$ enforcing on $v_{0|k}$: a) no bounds (solid line); b) hard bounds (dotted line); and, c) soft bounds (dashed lined).}
\label{fig:alltraj} 
\end{figure}
The drawback is a non negligible violation on the input constraint, as shown in Figure \ref{fig:allstatein}.

On the other hand, the  {MS-SMPC} with hard bounds on the first predicted input (case B) leads to a slower convergent trajectory, as depicted in Figure~\ref{fig:caseB0}, and necessitates of higher relaxations on the state constraints, as witnessed by the larger sequence of ellipsoidal sets. However, this approach ensures no input constraints violation (see Figure~\ref{fig:allstatein}).

Finally, the soft input strategy (case C), which is a trade-off between the other two, leads to a state trajectory that is faster than the one obtained with hard constraints, but with more input violations occurrences, as illustrated in Figure~\ref{fig:caseC0}. Moreover, as depicted in Figure~\ref{fig:allstatein}, the input constraints are violated in the first three steps but with lower values, with respect to the strategy of no bounds, leading to slower prediction and then higher state constraints violations probabilities.

The considerations above are corroborated also by the results depicted in Figure~\ref{fig:alltraj}, representing the realized trajectories obtained along a simulation horizon of 10 steps for the three strategies. It can be noticed that the tighter the constraints on the initial input are, the slower the trajectory convergence rate is. 

\begin{figure}
\begin{center}
%
%
\definecolor{mycolor1}{rgb}{0.00000,0.44700,0.74100}%
\definecolor{mycolor2}{rgb}{0.85000,0.32500,0.09800}%
\definecolor{mycolor3}{rgb}{0.92900,0.69400,0.12500}%
\begin{tikzpicture}

\begin{axis}[%
width=7.5cm,
height=5cm,
at={(0in,0in)},
scale only axis,
xmin=9095,
xmax=18005,
ymin=0,
ymax=300,
axis background/.style={fill=white},
legend style={legend cell align=left, align=left, draw=white!15!black}
]
\addplot[ybar interval, fill=mycolor1, fill opacity=0.6, draw=black, area legend] table[row sep=crcr] {%
x	y\\
9500	1\\
9600	15\\
9700	62\\
9800	168\\
9900	274\\
10000	228\\
10100	159\\
10200	77\\
10300	11\\
10400	5\\
10500	5\\
};
\addlegendentry{No bounds}

\addplot[ybar interval, fill=mycolor2, fill opacity=0.6, draw=black, area legend] table[row sep=crcr] {%
x	y\\
13100	1\\
13200	0\\
13300	0\\
13400	1\\
13500	1\\
13600	2\\
13700	2\\
13800	4\\
13900	2\\
14000	8\\
14100	12\\
14200	7\\
14300	13\\
14400	17\\
14500	17\\
14600	33\\
14700	33\\
14800	37\\
14900	55\\
15000	53\\
15100	53\\
15200	61\\
15300	61\\
15400	59\\
15500	57\\
15600	58\\
15700	53\\
15800	52\\
15900	46\\
16000	43\\
16100	28\\
16200	26\\
16300	17\\
16400	22\\
16500	17\\
16600	13\\
16700	8\\
16800	4\\
16900	10\\
17000	4\\
17100	3\\
17200	2\\
17300	1\\
17400	3\\
17500	1\\
17600	1\\
};
\addlegendentry{Hard bounds}

\addplot[ybar interval, fill=mycolor3, fill opacity=0.6, draw=black, area legend] table[row sep=crcr] {%
x	y\\
11000	1\\
11100	0\\
11200	3\\
11300	6\\
11400	12\\
11500	18\\
11600	41\\
11700	54\\
11800	93\\
11900	76\\
12000	122\\
12100	128\\
12200	111\\
12300	79\\
12400	86\\
12500	63\\
12600	47\\
12700	24\\
12800	12\\
12900	12\\
13000	4\\
13100	3\\
13200	3\\
13300	0\\
13400	2\\
13500	2\\
};
\addlegendentry{Soft bounds}

\end{axis}
\end{tikzpicture}%
\caption{Cost comparison over 1000 simulations with $N=10$ for $x_0 = (-40,\, 40)$ of the three input bounds strategies. Mean cost: no bounds $J_{MPC} = 10006$; hard bounds $J_{MPC}=15467$; soft bounds $J_{MPC}=12173$. \label{fig:hists4040}}
\end{center}
\end{figure}
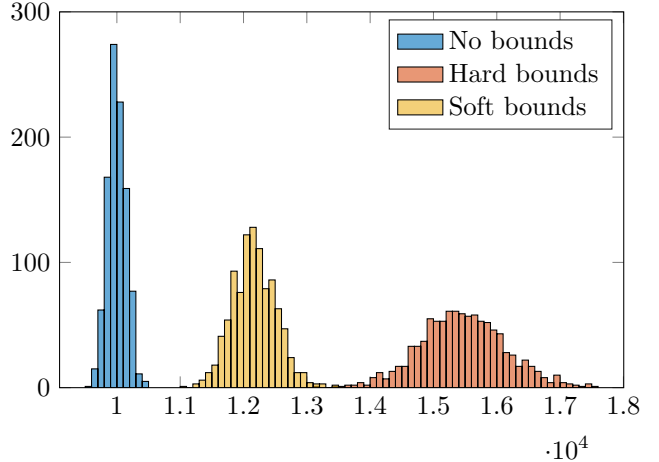

Moreover, the three strategies have been compared in terms of the following performance index
\begin{equation}\label{eq:costMPC}
J_{MPC} \doteq\,\, \sum_{k = 0}^{N-1} \left(\|x_k\|_Q^2+ \|u_k\|_R^2\right)
\end{equation}
over 1000 trajectories with $N=10$ and $x_0=(-40,\,40)$. The histograms of the resulting cost values are depicted in Figure~\ref{fig:hists4040}. It can be noticed that, the tighter are the bounds on the initial input $v_{0|k}$, the worst are the control performances, which is reasonable.

\subsection{Probability bounds comparison}
As illustrated in Section~\ref{sec:pr_state}, the constraints relaxation obtained by introducing $\bar{r}_{x}$ and $\bar{r}_{u}$ as free variables in the optimization problem \eqref{eq:SMPC_mixed} can also be interpreted as a relaxation in the probability levels of the chance constraints. In particular, by determining $\bar{\rho}_{\ell|x}$ as in \eqref{eq:barrxk} and $\bar{\rho}_{\ell|u}$ as in \eqref{eq:barruk} for a solution $z_{\ell|k}$ and $v_{\ell|k}$ with $\ell = \mathbb{N}_0^N$, probability bounds on the chance constraints satisfaction can be obtained for the specific measured state $x_k$. Table~\ref{tab:1} provides the probability bound values 
\begin{equation}\label{eq:rho}
p_\ell(j) = \big(p_{\ell|x}(j), \, p_{\ell|u}(j)\big) = \Big( \chi_n^2 \big(\bar{\rho}_{\ell|x}^2(j) \big) , \ \chi_n^2 \big(\bar{\rho}_{\ell|u}^2(j)\big) \Big)
\end{equation}
with $\bar{\rho}_{\ell|x}(j)$ and $\bar{\rho}_{\ell|u}(j)$ computed as in \eqref{eq:barrxk}--\eqref{eq:barruk} for each input bound strategy, i.e., $j = A, B, C$, and for all $\ell = \mathbb{N}_1^{10}$, starting from the optimal prediction pair, $z_{\ell|k}$ and $v_{\ell|k}$, computed at $k = 0$. Moreover, $N_{sim} = 1000$ simulations have been run for the different strategies and the occurrence of violations of the constraints $x_k \in \Ec_{W_x}\!(r_{x})$ and $u_k \in \Ec_{W_u}\!(r_{u})$ are registered to obtain the relative frequency of constraints satisfaction given by
\begin{align}
f_\ell(j) & = \big(f_{\ell|x}(j), \, f_{\ell|u}(j)\big) \nonumber\\
& = \left( \frac{N_{sim} - n_{\ell|x}(j)}{N_{sim}}, \ \frac{N_{sim} - n_{\ell|u}(j)}{N_{sim}} \right) \label{eq:f}
\end{align}
where $n_{\ell|x}(j)$ and $n_{\ell|u}(j)$ are the number of violations at time $\ell$ of the state and input constraints, respectively, with $j = A,B,C$. The values of $f_\ell(j)$ are reported in the right part of Table~\ref{tab:1}. It is worth noting that bounds on the future constraints satisfaction, based on the optimal nominal states and inputs obtained at time $k = 0$, are reasonably accurate guesses. 

Moreover, we can note how the constraints on the first input $v_{0|k}$ affect the violation probabilities of the MS-SMPC. Larger freedom on the first input selection leads to a more aggressive MPC action for driving the state towards the feasibility region with no relaxation, hence to bigger violations on the input constraints (see Figure~\ref{fig:allstatein}) but less frequent violation on the future state constraints, which is also reasonable.

\begin{table}[!ht]
\centering
\setlength\tabcolsep{3pt} 
\begin{tabular}{|c|ccc|ccc|}
 \hline \hline
$\ell$ &  $p_\ell(A)$ & $p_\ell(B)$ & $p_\ell(C)$ & $f_\ell(A)$ &  $f_\ell(B)$ & $f_\ell(C)$\\
\hline
1  & (1,0.90)	& (0,0) 		& (0,0) 	& (1,0)	  	& (0,0)		& (0,0)     \\
2  & (1,1) 		& (0, 0) 		& (0,0) 	& (1,1)    	& (0,0)     & (0,0)   	\\
3  & (1,1) 		& (0,0.92) 		& (0.73,1) 	& (1,1)    	& (0,0.89)  & (0.15,0.92)  \\
4  & (1,1) 		& (0.99, 1) 	& (1,1) 	& (1,1)    	& (0.01,1)  & (1,1)     \\
5  & (1,1) 		& (1,1)		  	& (1,1)		& (1,1)    	& (0.96,1)  & (1,1)     \\
6  & (1,1) 		& (1,1)   		& (1,1)		& (1,1)    	& (1,1)     & (1,1)     \\
7  & (1,1)		& (1,1)   		& (1,1)		& (1,1)    	& (1,1)     & (1,1)     \\
8  & (1,1)		& (1,1)   		& (1,1)		& (1,1)    	& (1,1)     & (1,1)     \\
9  & (1,1)		& (1,1)   		& (1,1)		& (1,1)    	& (1,1)     & (1,1)     \\
10 & (1,1)    	& (1,1)       	& (1,1)     & (1,1)		& (1,1)   	& (1,1)\\
\hline \hline
\end{tabular}
\caption{Probability of state and input constraints satisfaction within the predicted bounds $p_\ell(j)$ defined in \eqref{eq:rho} with $\bar{\rho}_{\ell|x}(j)$ and 
$\bar{\rho}_{\ell|u}(j)$ as in \eqref{eq:barrxk} and \eqref{eq:barruk} for $j = A, B, C$, and  relative frequencies $f_\ell(j)$ defined in \eqref{eq:f} of both state and input constraints satisfaction along the trajectories for MS-SMPC over $N_{sim} = 1000$ tests each, for $\ell \in \N_1^{10}$ and $x_0 = (-40, \, 40)$. \label{tab:1}
}
\end{table}

\begin{figure}
\begin{center}
%
%
\definecolor{mycolor1}{rgb}{0.00000,0.44700,0.74100}%
\definecolor{mycolor2}{rgb}{0.85000,0.32500,0.09800}%
\begin{tikzpicture}

\begin{axis}[%
width=7.5cm,
height=5cm,
at={(0in,0in)},
scale only axis,
xmin=2736.75,
xmax=3248.25,
ymin=0,
ymax=120,
axis background/.style={fill=white},
legend style={legend cell align=left, align=left, draw=white!15!black}
]
\addplot[ybar interval, fill=mycolor1, fill opacity=0.6, draw=black, area legend] table[row sep=crcr] {%
x	y\\
2760	2\\
2775	2\\
2790	3\\
2805	13\\
2820	17\\
2835	17\\
2850	36\\
2865	44\\
2880	55\\
2895	76\\
2910	83\\
2925	97\\
2940	81\\
2955	95\\
2970	86\\
2985	66\\
3000	55\\
3015	45\\
3030	39\\
3045	33\\
3060	18\\
3075	11\\
3090	6\\
3105	7\\
3120	5\\
3135	6\\
3150	0\\
3165	2\\
3180	2\\
};
\addlegendentry{MS-SMPC}

\addplot[ybar interval, fill=mycolor2, fill opacity=0.6, draw=black, area legend] table[row sep=crcr] {%
x	y\\
2775	5\\
2790	3\\
2805	7\\
2820	10\\
2835	29\\
2850	30\\
2865	40\\
2880	52\\
2895	81\\
2910	64\\
2925	110\\
2940	98\\
2955	100\\
2970	72\\
2985	66\\
3000	59\\
3015	55\\
3030	47\\
3045	30\\
3060	16\\
3075	10\\
3090	7\\
3105	5\\
3120	1\\
3135	1\\
3150	0\\
3165	1\\
3180	0\\
3195	0\\
3210	1\\
3225	1\\
};
\addlegendentry{IS-SMPC}

\end{axis}
\end{tikzpicture}%
\caption{Cost comparison over 1000 simulations with $N=10$ for $x(0) = (-30, 0)$. Mean cost: for the new  {MS-SMPC} (with no initial bound) $J_{MPC}=2952$, for the  {IS-SMPC} $J_{MPC}=2953$, with almost unitary ratio. \label{fig:3000}}
\end{center}
\end{figure}
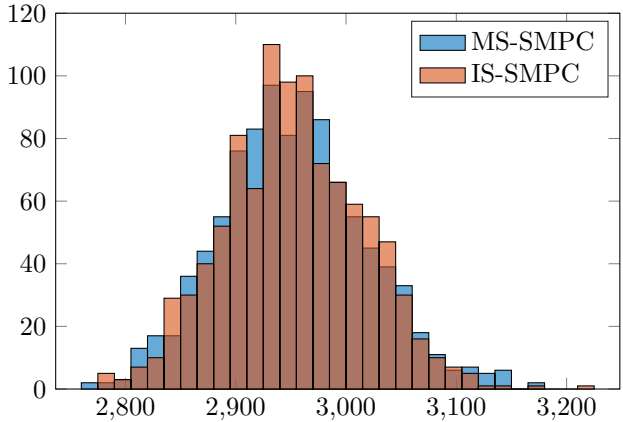
    
\subsection{Comparison of  {IS-SMPC} and  {MS-SMPC}}
In this second set of simulations, we consider a case where both IS-SMPC and MC-SMPC are feasible. We compare the  {MS-SMPC} (with no bounds on in the initial input, case A) with the  {IS-SMPC} in terms of the performance index $J_{MPC}$ defined in \eqref{eq:costMPC}. For the first case study, we consider as initial condition $x(0) = (-30,0)$, which is inside the feasibility region and far from its boundaries. For each SMPC scheme, 1000 simulations have been run and we can observe in Figure~\ref{fig:3000} that both control schemes lead, in practice, to the same unconstrained optimization problem and, consequently, to the same solutions. Indeed, the histograms of the cost function realized along the generated trajectories and depicted in Figure~\ref{fig:3000} are almost overlapped and the mean value of the performance index almost the same.

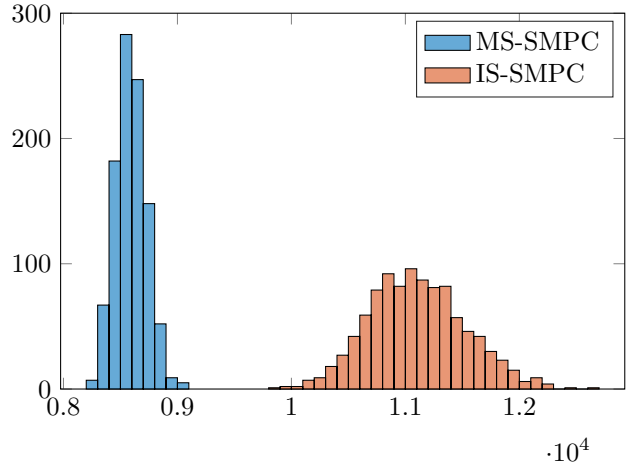
\begin{figure}
\begin{center}
%
%
\definecolor{mycolor1}{rgb}{0.00000,0.44700,0.74100}%
\definecolor{mycolor2}{rgb}{0.85000,0.32500,0.09800}%
\begin{tikzpicture}

\begin{axis}[%
width=7.5cm,
height=5cm,
at={(0in,0in)},
scale only axis,
xmin=7975,
xmax=12925,
ymin=0,
ymax=300,
axis background/.style={fill=white},
legend style={legend cell align=left, align=left, draw=white!15!black}
]
\addplot[ybar interval, fill=mycolor1, fill opacity=0.6, draw=black, area legend] table[row sep=crcr] {%
x	y\\
8200	7\\
8300	67\\
8400	182\\
8500	283\\
8600	247\\
8700	148\\
8800	52\\
8900	9\\
9000	5\\
9100	5\\
};
\addlegendentry{MS-SMPC}

\addplot[ybar interval, fill=mycolor2, fill opacity=0.6, draw=black, area legend] table[row sep=crcr] {%
x	y\\
9800	1\\
9900	2\\
10000	2\\
10100	7\\
10200	9\\
10300	18\\
10400	27\\
10500	42\\
10600	59\\
10700	79\\
10800	92\\
10900	82\\
11000	96\\
11100	87\\
11200	81\\
11300	82\\
11400	57\\
11500	46\\
11600	42\\
11700	30\\
11800	23\\
11900	15\\
12000	6\\
12100	9\\
12200	4\\
12300	0\\
12400	1\\
12500	0\\
12600	1\\
12700	1\\
};
\addlegendentry{IS-SMPC}

\end{axis}
\end{tikzpicture}%
\caption{Cost comparison over 1000 simulations of 10 steps for $x(0) = (-40, 37)$, close to the  {IS-SMPC} feasibility bounds. Mean cost: for the  {MS-SMPC} (with no initial bound) $8593$, for the IF-SMPC $11105$, with ratio of around $0.77$. \label{fig:4037}}
\end{center}
\end{figure}

On the other hand, for initial conditions closer to the boundary of the feasibility region, we can observe quite distinct behaviours of the two SMPC schemes. In particular, running 1000 simulations starting from $x(0) = (-40,37)$, we can notice a significantly different effect of the constraints on the realized trajectories, as shown in Figure~\ref{fig:4037}. In this second case study, the histograms representing the cost $J_{MPC}$ for the obtained realizations show that the performance for the  {MS-SMPC} substantially outperforms the one granted by the  {IS-SMPC}, being the mean of the former less than $80\%$ of the latter one.

\section{Conclusion and future works}
\label{sec:concl}

Motivated by the difficulty of guaranteeing recursive feasibility of SMPC problems in the presence of unbounded stochastic  uncertainties, especially when one wants to exploit the knowledge of the measured state in implementing the
feedback strategy, in this paper we
introduced a novel definition of recursive feasibility. Namely, we allow for a relaxation of the constraints so that the ensuing optimization problem remains feasible, but we require that such relaxation is minimal, and more importantly, it has certain probabilistic properties. In parallel, we developed a novel SMPC approach guaranteeing these new properties. We prove, also by numerical simulations, that this new approach allows to tackle problems not affordable by classical open-loop strategies, and that it outperforms these latter when both are feasible. In our next studies, we aim at extending the proposed philosophy to other SMPC frameworks. For instance, it would be interesting to investigate how the introduced concepts may extend to problems involving
\textit{mission-wide constraint}, first introduced in \cite{Lew2019ChanceConstrainedOA}.

\appendix
\section*{Appendix}

\section{Proof of Theorem \ref{th:rec_feas}}\label{app:proof3}
\begin{proof}
The proof follows the standard procedure to demonstrate recursive feasibility with a shifting in the optimal solution $\textbf{v}^\star_k=[v^\star_{0|k},\ldots,v^\star_{N-1|k}]$ of optimization problem \eqref{eq:BasicSMPC} at time $k$ and enforcing the terminal controller $v_{N|k}=Kz^\star_{N|k}$ in the final prediction step, i.e., $v^\star_{k+1}=\{v^\star_{1|k},\ldots,v^\star_{N-1|k},Kz^\star_{N|k}\}$. Then, we have that the first $N-1$ entries fulfill input constraints again, and furthermore according to Theorem~\ref{th:term} we see that the final entry $\bar v_{N-1|k+1}=Kz^\star_{N|k}$ belong to the terminal region. Then, the resulting trajectory for the nominal state is $\textbf{z}^\star=[z^\star_{1|k},\ldots,z^\star_{N|k},A_Kz^\star_{N|k}]$. Due to an analogue argument, this satisfies the nominal state constraints $\forall \ell$, and $A_Kz^\star_{N|k}$ belongs to the terminal region by constructions, according to Theorem~\ref{th:term}. Moreover, due to the specific choice of constraint tightening using PRS, and the linear evolution of the closed-loop error \eqref{eq:sys_e}, we can easily prove the chance constraint satisfaction in closed-loop. Following similar procedures to those in \cite{hewing2020recursively}, we have that by definition $\Pr{e_{\ell|k}\in|x_0}\geq 1-\phi_\varepsilon(\rho)$ for all $\ell\in\mathbb{N}_0^{N}$ and if Problem \eqref{eq:BasicSMPC} is feasible, consequently we have that $\Pr{x_{0|k}\in\mathbb{X}\,|x_0}\geq 1-\phi_\varepsilon(\rho)$ holds. The same argument holds also for input constraints.
\end{proof}

\section{Proof of Theorem \ref{th:feasibility}}\label{app:proof6}

\subsection{Proof of condition \textit{(i)}}
Condition {\it (i)} is straightforward since for every $x_k\in\mathbb{R}^n$ and every bounded   $v_{\ell|k}$ with $\ell \in \N_0^{N-1}$, also the sequence $z_{\ell|k}$ stays bounded. Hence, for every $x_k\in\mathbb{R}^n$ there exist $\bar{r}^\star_x$ and $\bar{r}^\star_u$ such that  \eqref{eq:prabar}-\eqref{eq:prcbar2} hold.

\subsection{Proof of condition \textit{(ii)}}
Moving to condition {\it (ii)}, we note that $x_{k+1}$ is a random variable with mean $\E{x_{k+1}} = A x_k + B v_{0|k}^\star(x_k)$ and covariance $\Gamma_w$, being $x_{k+1} = \E{x_{k+1}} + w_k$. Hence, also $\textbf{z}^\star(x_{k+1})$, $\textbf{v}^\star(x_{k+1})$, $\bar{r}_x^\star(x_{k+1})$, and $\bar{r}_u^\star(x_{k+1})$ are random variables. Moreover, since their values are defined for every realization of $w_k$, their expected value with respect to the distribution of $w_k$ conditioned to $x_k$ can be computed.

\subsubsection{Proof of \eqref{eq:condru}}
We first prove condition \eqref{eq:condru}. 
Consider the deterministic tightened constraints on the nominal input \eqref{eq:prbbar}, and define the candidate nominal input solution as follows
\begin{equation}\label{eq:th1v}
v_{\ell|k+1} = v_{\ell+1|k}^\star + K A_K^{\ell} w_k, \quad \forall \ell \in \N_0^{N-2},
\end{equation}
where the dependencies on $x_k$ and $x_{k+1}$ is left implicit to ease the notation, recalling that the realization of the random variable $w_k$ is known at time $k+1$.
Moreover, we can notice that the nominal input law \eqref{eq:th1v} is such that the predicted input $u_{\ell+1|k}^\star$ at time $k$ and the candidate one $u_{\ell|k+1}$  at time $k+1$ are the same, namely
 {
\begin{align}
    u^\star_{\ell+1|k} & = v_{\ell+1|k}^\star + K e_{\ell+1|k} \nonumber\\ 
    & = v_{\ell+1|k}^\star + K \sum_{i=0}^\ell A_K^{\ell-i} w_{k+i},   \label{eq:u1}\\
u_{\ell|k+1} & = v_{\ell|k+1} + K e_{\ell|k+1} \nonumber \\
& = v_{\ell|k+1} + K \sum_{i=1}^{\ell} A_K^{\ell-i} w_{k+i}, \label{eq:u2}
\end{align}
}
for $\ell \in \N_0^{N-2}$. Indeed, \eqref{eq:u1} and \eqref{eq:u2} are equal if and only if \eqref{eq:th1v} holds. It will be proved next that, since the deterministic constraints \eqref{eq:prbbar} is satisfied by $v_{\ell+1|k}$ with $\bar{r}_u = \bar{r}_u^\star$, then it also holds in expectation at $k+1$, i.e., $\E{(v_{\ell|k+1})^\top W_u^{-1}v_{\ell|k+1}} \leq (\bar{r}_u^\star - \rho (1 - \lambda^{\ell}))^2$, for all $\ell \in \N_0^{N-2}$.

Let us define the random variable $y = W_u^{-1/2}v_{\ell|k+1}$. From \eqref{eq:th1v}, we get $ y = W_u^{-1/2}(v_{\ell+1|k}^\star + K A_K^{\ell} w_k)$, and consequently $\E{y} = W_u^{-1/2}v_{\ell|k}^\star$ and $\cov{y} = \cov{ W_u^{-1/2} A_K^\ell w_k}$. Knowing that, given a random vector $y$, we have that $\E{y y^\top} = \cov{y} + \E{y}\E{y}^\top$, from \eqref{eq:reachLMI_1} and \eqref{eq:reachLMI_2} we get
%
\begin{align*}
& \hspace{-0.5cm} \E{yy^\top} = \cov{W_u^{-1/2}K A_K^{\ell} w_k} \\
& \quad + (W_u^{-1/2}v_{\ell|k}^\star)(W_u^{-1/2}v_{\ell|k}^\star)^\top\\
& = \E{W_u^{-1/2} K A_K^{\ell} w_kw_k^\top (A_K^{\ell})^\top K^\top W_u^{-1/2} }\\
& \quad + W_u^{-1/2}v_{\ell+1|k}^\star (v_{\ell+1|k}^\star)^\top W_u^{-1/2}\\
& = W_u^{-1/2} K A_K^{\ell} \Gamma_w (A_K^{\ell})^\top K^\top W_u^{-1/2} \\
& \quad + W_u^{-1/2}v_{\ell+1|k}^\star (v_{\ell+1|k}^\star)^\top W_u^{-1/2}\\
& \preceq \lambda^{2\ell}(1-\lambda)^2  W_u^{-1/2} K W_x K^\top W_u^{-1/2} \\
& \quad + W_u^{-1/2}v_{\ell+1|k}^\star (v_{\ell+1|k}^\star)^\top W_u^{-1/2}\\
& \preceq \lambda^{2\ell}(1-\lambda)^2 \mathbb{I}_n + W_u^{-1/2}v_{\ell+1|k}^\star (v_{\ell+1|k}^\star)^\top W_u^{-1/2}
\end{align*}
where the last inequality follows from the fact that 
\begin{equation*}
 W_u^{-1/2} K W_x K^\top W_u^{-1/2} \preceq \mathbb{I}_n
\end{equation*}
which holds from \eqref{eq:WuWx}. Then, we can compute the expected value of $\E{v_{\ell|k+1}^\top  W_u^{-1}v_{\ell|k+1}}$ relying on the property of random variable for which we have that $\E{y^\top y}=\tr{\E{yy^\top}}$. Hence, we obtain
\begin{align*}
\tr{\E{yy^\top}} & \leq \tr{W_u^{-1/2}v_{\ell+1|k}^\star (v_{\ell+1|k}^\star)^\top W_u^{-1/2}} \\
& \quad +\tr{(1-\lambda)^2 \lambda^{2\ell} \mathbb{I}_n}\\
& = (v_{\ell+1|k}^\star)^\top W_u^{-1} v_{\ell+1|k}^\star + (1-\lambda)^2 \lambda^{2\ell} n\\
& \leq (\bar{r}_u - \rho (1 - \lambda^{\ell+1}))^2+(1-\lambda)^2 \lambda^{2\ell} n 
\end{align*}
where the last inequality follows from the fact that $v_{\ell+1|k}^\star$ satisfies \eqref{eq:prbbar}. Now, we look for a condition over the parameter $\rho$ which guarantees that the constraint \eqref{eq:prbbar} is satisfied in expectation by $v_{\ell|k+1}$ with the same $\bar{r}_u$, i.e.,
$$\E{v_{\ell|k+1}^\top W_u^{-1} v_{\ell|k+1}}\leq (\bar{r}_u - \rho (1 - \lambda^{\ell}))^2,$$
which corresponds to define a condition on $\rho$ for which
\begin{equation}\label{eq:th1eta}
(\bar{r}_u - \rho (1 - \lambda^{\ell+1})^2 + (1-\lambda)^2 \lambda^{2\ell} n\leq (\bar{r}_u - \rho (1 - \lambda^{\ell}))^2
\end{equation}
holds. Specifically, employing \eqref{eq:rubar} and after some manipulation, we can prove that   
\eqref{eq:th1eta} holds if  
\eqref{eq:th1boundrho} is satisfied, which also provides a condition on the violation probability $\phi_\varepsilon(\rho)$.

The analysis above holds for $\ell \in \N_0^{N-2}$, for which the nominal input $v_{\ell|k+1}$ can be defined as a function of $v_{\ell+1|k}^\star$ \eqref{eq:th1v}. Now, let us consider the nominal input defined at time $\ell = N-1$ as
\begin{align}\label{eq:th1vN}
v_{N-1|k+1} = K z_{N|k}^\star + K A_K^{N-1} w_k, 
\end{align}
given \eqref{eq:th1v} and knowing that $v_{N|k}^\star= K z_{N|k}^\star$.

By defining $y = W_u^{-1/2} v_{N-1|k+1} = W_u^{-1/2} (K z_{N|k} + K A_K^{N-1} w_k)$, the expected value of $\E{y^\top y}$ can be bounded to prove that \eqref{eq:condru} holds also for the last element of the nominal input sequence solving the problem \eqref{eq:SMPC_barr} at $k+1$. Following a similar procedure as before and applying \eqref{eq:reachLMI_1}, \eqref{eq:reachLMI_2}, \eqref{eq:WuWx}, and \eqref{eq:prcbar2}, we have
\begin{align*}
& \hspace{-0.3cm} \E{yy^\top} = \E{W_u^{-1/2} K A_K^{\ell} w_kw_k^\top (A_K^{\ell})^\top K^\top W_u^{-1/2} }\\
& \quad + W_u^{-1/2} K z_{N|k} z_{N|k}^\top K^\top W_u^{-1/2}\\
& \preceq (1-\lambda)^2 \lambda^{2N-2} \mathbb{I}_n+ W_u^{-1/2}  K z_{N|k} z_{N|k}^\top K^\top W_u^{-1/2}
\end{align*}
from which we get
\begin{equation}\label{eq:th1temp1}
\E{y^\top y} \leq (1-\lambda)^2 \lambda^{2N-2} n + (\bar{r}_u - \rho (1 - \lambda^{N})^2.
\end{equation}
Then if the following condition holds
$$(\bar{r}_u - \rho (1 - \lambda^{N})^2+(1-\lambda)^2 \lambda^{2N-2} n \leq (\bar{r}_u - \rho (1 - \lambda^{N-1})^2,$$
then the expectation of $v_{N-1|k+1}^\top W_u^{-1}v_{N-1|k+1}$ satisfies the constraint \eqref{eq:prbbar} with the same value of $\bar{r}_u$, and it can be proved that such condition holds if \eqref{eq:th1boundrho} is satisfied.

\subsubsection{Proof of \eqref{eq:condrx}}
Now, let us prove condition \eqref{eq:condrx} related to the constraints on the predicted nominal states. Following an analogous reasoning, it can be proved that $z_{\ell+1|k}$ satisfying \eqref{eq:prabar} implies its satisfaction in expectation also for $z_{\ell|k+1}$. Notice that 
\begin{equation*}
x_{k+1} = z_{0|k+1} = Ax_k + B v_{0|k}^\star + w_k = z_{1|k} + w_k.
\end{equation*}
Moreover, the nominal trajectory at $k+1$, for a given $w_k$ and with nominal control \eqref{eq:th1v} and \eqref{eq:th1vN} is given by 
\begin{align}
z_{1|k+1} & = A z_{0|k+1} + B v_{1|k}^\star + B K w_k \nonumber\\
& = A z_{1|k} + A w_k + B v_{1|k}^\star + B K w_k \nonumber\\
& = z_{2|k} + A_K w_k  \nonumber\\
 \vdots \quad &\nonumber\\
z_{\ell|k+1} & = z_{\ell+1|k} + A_K^{\ell} w_k, \qquad \forall \ell \in \N_0^{N-1}. \label{eq:th1temp2}
\end{align}
Proceeding as above, it can be proved that the random variable $y = W_x^{-1/2}(z_{\ell|k+1}) = W_x^{-1/2}(z_{\ell+1|k} + A_K^\ell w_k)$ is such that 
\begin{equation}\label{eq:th1condz}
\E{y^\top y} \leq (\bar{r}_x^\star - \rho(1-\lambda^\ell))^2
\end{equation}
if condition \eqref{eq:prabar} holds for $z_{\ell+1|k}$. In fact, since $\E{y} = W_x^{-1/2}z_{\ell+1|k}$ and $\cov{y} = \cov{ W_x^{-1/2} A_K^\ell w_k}$, we have
\begin{align*}
& \hspace{-0.3cm} \E{yy^\top} = \E{W_x^{-1/2} A_K^{\ell} w_kw_k^\top(A_K^{\ell})^\top W_x^{-1/2} }\\
& \quad + W_x^{-1/2}z_{\ell+1|k} (z_{\ell+1|k})^\top W_z^{-1/2}\\
& = W_x^{-1/2} A_K^{\ell} \Gamma_w (A_K^{\ell})^\top W_x^{-1/2} \\
& \quad + W_x^{-1/2}z_{\ell+1|k} (z_{\ell+1|k})^\top W_x^{-1/2}\\
& \preceq (1-\lambda)^2 \lambda^{2\ell} \mathbb{I}_n + W_x^{-1/2}z_{\ell+1|k} (z_{\ell+1|k})^\top W_x^{-1/2},
\end{align*}
from which we get
\begin{align*}
\E{y^\top y} & \leq \tr{(1-\lambda)^2 \lambda^{2\ell} \mathbb{I}_n} \\ 
& \quad + \tr{W_x^{-1/2}z_{\ell+1|k} (z_{\ell+1|k})^\top W_z^{-1/2}} \\
& = (1-\lambda)^2 \lambda^{2\ell} n + (z_{\ell+1|k})^\top W_x^{-1} z_{\ell+1|k} \\
& \leq (1-\lambda)^2 \lambda^{2\ell} n + (\bar{r}_x - \rho (1 - \lambda^{\ell+1}))^2,
\end{align*}
where the last inequality is holds since $z_{\ell+1|k}$ satisfies \eqref{eq:prabar} with the specific value $\bar{r}_x = \bar{r}_x^\star$. Then, using \eqref{eq:rxbar},  it can be proved that the condition \eqref{eq:th1boundrho} on $\rho$ is sufficient for \eqref{eq:th1condz} to hold.

Finally, let us consider the nominal terminal state at $k+1$, which can be rewritten as follows combining \eqref{eq:th1temp2} and \eqref{eq:th1vN}, i.e.,  
\begin{align*}
z_{N|k+1} & = A z_{N-1|k+1} + B v_{N-1|k+1}\\
& = A (z_{N|k} + A_K^{N-1} w_k) + BK(z_{N|k} + A_K^{N-1}w_k)\\
& = A_K z_{N|k} + A_K^{N}w_k.
\end{align*}
From the same reasoning applied but selecting as random variable $y = W_x^{-1/2}z_{N|k+1}$, one has  
\begin{align}
\E{y^\top y} & \leq (1-\lambda)^2 \lambda^{2N} n + \lambda^2 (\bar{r}_x - \rho (1 - \lambda^{N}))^2. \label{eq:th1temp3}
\end{align}
Hence, we can prove that \eqref{eq:prcbar} is satisfied in expectation at $k+1$ with the same $\bar{r}_x$ if 
\begin{equation}\label{eq:th1temp3b}
(1-\lambda)^2 \lambda^{2N} n + \lambda^2 (\bar{r}_x - \rho (1 - \lambda^{N}))^2 \leq (\bar{r}_x - \rho (1 - \lambda^{N}))^2
\end{equation}
which holds if 
\begin{equation*}
\lambda^N\left(\sqrt{n\frac{1-\lambda}{1+\lambda}} - \rho \right) + \rho \leq \bar{r}_x,
\end{equation*}
that is satisfied if \eqref{eq:th1boundrho} holds, which renders the first term non-positive, and knowing that $\rho \leq \bar{r}_x$ from \eqref{eq:rxbar}.

The last constraint to be proved to hold for the expected solution of problem \eqref{eq:SMPC_barr} at time $k+1$ is \eqref{eq:prcbar2}. Analogously to  \eqref{eq:th1temp3} and from \eqref{eq:prcbar2} holding with $z_{N|k}$, it follows that \eqref{eq:prcbar2} holds in expectation if 
\begin{equation*}\label{eq:th1temp4}
\lambda^2 (\bar{r}_u - \rho (1 - \lambda^{N}))^2 + (1-\lambda)^2 \lambda^{2N} n\leq (\bar{r}_u - \rho (1 - \lambda^{N}))^2,
\end{equation*}
that is satisfied for 
\begin{equation*}
\lambda^N\left(\sqrt{n\frac{1-\lambda}{1+\lambda}} - \rho \right) + \rho \leq \bar{r}_u.
\end{equation*}
which holds given condition \eqref{eq:th1boundrho} and constraint \eqref{eq:rubar}.

\subsubsection{Proof of \eqref{eq:condr}}
Finally, satisfaction of \eqref{eq:condru} and \eqref{eq:condrx} implies \eqref{eq:condr}.

\subsection{Proof of condition \textit{(iii)}}
{Last, to demonstrate condition {\it (iii)}, it is sufficient to note that dividing \eqref{eq:condru} and \eqref{eq:condrx} by $r_u$ and $r_x$, respectively, yields to MSRF condition \eqref{eq:RCFE_def} with $\gamma_u^\star(x)=\bar{r}_u^\star(x)/r_u$ and $\gamma_x^\star(x)=\bar{r}_x^\star(x)/r_x$, i.e., condition \eqref{eq:gammar}. Moreover, since feasibility at time $k$ with \eqref{eq:gammar} also implies ISRF conditioned to $x_k$ (see Theorem~\ref{th:rec_feas}), then MSRF in expectation and conditions \eqref{eq:constr_xk} hold}.

\section{Proof of Theorem \ref{th:balanced_cost}}\label{app:proof5}
\begin{proof}
Let $\mathbb{E}\{J_{r}^\star(x_{k+1})\,|\,x_k\}$ be the expected optimal value at time $k+1$ of Problem \eqref{eq:SMPC_mixed}, conditioned on the state $x_k$ and on the feasibility of the candidate solution $\bar v_{\ell|k+1}=v^\star_{\ell+1|k}$. 
We have
\begin{equation*}
    \begin{aligned}
        &\quad \,\,\mathbb{E}\{J_{r}^\star(x_{k+1})\,|\,x_k\}-J_{r}^\star(x_k)\\
        &\leq \mathbb{E}\Bigg\{\sum_{\ell=0}^{N-1}\Big(\|\bar z_{\ell|k+1}\|_Q^2+\|\bar v_{\ell|k+1}\|_R^2\Big)+V_f(\bar z_{N|k+1})\Bigg\}\\
        &\quad -\sum_{\ell=0}^{N-1}\Big(\|z^\star_{\ell|k}\|_Q^2+\|v^\star_{\ell|k}\|_R^2\Big)+V_f(z^\star_{N|k})\\
        &\quad +\mathbb{E}\{\Delta r^\star(x_{k+1})\,|\, x_k\}-\Delta r^\star(x_{k})\\
        & =\mathbb{E}\Bigg\{\|z^\star_{N|k}\|_Q^2+\|v^\star_{N|k}\|_R^2+V_f(z^\star_{N+1|k})\Bigg\}\\
        &\quad -\mathbb{E}\Bigg\{\|x^\star_{0|k}\|_Q^2+\|u^\star_{0|k}\|_R^2+V_f(z^\star_{N|k})\Bigg\}\\
       & \quad +\mathbb{E}\{\Delta r^\star(x_{k+1})\,|\, x_k\}-\Delta r^\star(x_{k})\\
       & \leq \mathbb{E}\Bigg\{\|z^\star_{N|k}\|_{Q+K^\top RK}^2+V_f(z^\star_{N+1|k})-V_f(z^\star_{N|k})\Bigg\}\\
       & \quad -\|z^\star_{0|k}\|_Q^2-\|v^\star_{0|k}\|_R^2 \leq-\|x_k\|_Q^2-\|u^\star_{0|k}\|_R^2,
    \end{aligned}
\end{equation*}
having applied the feedback law $v_{\ell|k}=Kz_{\ell|k}$ for all $\ell\geq N$ and the condition in \eqref{eq:condr}. Hence, we have obtained an upper bound on the cost decrease in expectation which depends only on $x_k$ and $u_{0|k}^\star$.
\begin{equation*}
\mathbb{E}\{J_{r}^\star(x_{k+1})\,|\,x_k\}-J_{r}^\star(x_k)\leq -\|x_k\|_Q^2-\|u^\star_{0|k}\|_R^2.
\end{equation*}
\end{proof}

\bibliographystyle{IEEEtran}
\bibliography{SMPC}

\end{document}